\documentclass[letterpaper,twocolumn,10pt]{article}
\usepackage{usenix2019,epsfig,endnotes}
\usepackage{enumerate}
\usepackage{multirow}
\usepackage{makecell}
\usepackage{url}
\usepackage{float}
\usepackage{array,graphicx}
\usepackage{booktabs}
\usepackage{hyperref}

\newcommand*\rot{\rotatebox{90}}

\newcolumntype{C}[1]{>{\centering\let\newline\\\arraybackslash\hspace{0pt}}m{#1}}

\newtheorem{definition}{Definition}

\usepackage{listings, xcolor}

\definecolor{verylightgray}{rgb}{.97,.97,.97}

\lstdefinelanguage{Solidity}{
	keywords=[1]{anonymous, assembly, assert, balance, break, call, callcode, case, catch, class, constant, continue, contract, debugger, default, delegatecall, delete, do, else, emit, event, export, external, false, finally, for, function, gas, if, implements, import, in, indexed, instanceof, interface, internal, is, length, library, log0, log1, log2, log3, log4, memory, modifier, new, payable, pragma, private, protected, public, pure, push, require, return, returns, revert, selfdestruct, send, storage, struct, suicide, super, switch, then, this, throw, transfer, true, try, typeof, using, value, view, while, with, addmod, ecrecover, keccak256, mulmod, ripemd160, sha256, sha3}, 
	keywordstyle=[1]\color{blue}\bfseries,
	keywords=[2]{address, bool, byte, bytes, bytes1, bytes2, bytes3, bytes4, bytes5, bytes6, bytes7, bytes8, bytes9, bytes10, bytes11, bytes12, bytes13, bytes14, bytes15, bytes16, bytes17, bytes18, bytes19, bytes20, bytes21, bytes22, bytes23, bytes24, bytes25, bytes26, bytes27, bytes28, bytes29, bytes30, bytes31, bytes32, enum, int, int8, int16, int24, int32, int40, int48, int56, int64, int72, int80, int88, int96, int104, int112, int120, int128, int136, int144, int152, int160, int168, int176, int184, int192, int200, int208, int216, int224, int232, int240, int248, int256, mapping, string, uint, uint8, uint16, uint24, uint32, uint40, uint48, uint56, uint64, uint72, uint80, uint88, uint96, uint104, uint112, uint120, uint128, uint136, uint144, uint152, uint160, uint168, uint176, uint184, uint192, uint200, uint208, uint216, uint224, uint232, uint240, uint248, uint256, var, void, ether, finney, szabo, wei, days, hours, minutes, seconds, weeks, years},	
	keywordstyle=[2]\color{teal}\bfseries,
	keywords=[3]{block, blockhash, coinbase, difficulty, gaslimit, number, timestamp, msg, data, gas, sender, sig, value, now, tx, gasprice, origin},	
	keywordstyle=[3]\color{violet}\bfseries,
	identifierstyle=\color{black},
	sensitive=false,
	comment=[l]{//},
	morecomment=[s]{/*}{*/},
	commentstyle=\color{gray}\ttfamily,
	stringstyle=\color{red}\ttfamily,
	morestring=[b]',
	morestring=[b]"
}

\begin{document}

\date{}

\title{\Large \bf The Art of The Scam: \\ Demystifying Honeypots in Ethereum Smart Contracts}


\author{
{\rm Christof Ferreira Torres}\\
SnT, University of Luxembourg
\and
{\rm Mathis Steichen}\\
SnT, University of Luxembourg
\and
{\rm Radu State}\\
SnT, University of Luxembourg
} 

\maketitle

\thispagestyle{empty}

\subsection*{Abstract}



Modern blockchains, such as Ethereum, enable the execution of so-called \emph{smart contracts} -- programs that are executed across a decentralised network of nodes.
As smart contracts become more popular and carry more value, they become more of an interesting target for attackers.
In the past few years, several smart contracts have been exploited by attackers.
However, a new trend towards a more proactive approach seems to be on the rise, where attackers do not search for vulnerable contracts anymore. 
Instead, they try to lure their victims into traps by deploying seemingly vulnerable contracts that contain hidden traps.
This new type of contracts is commonly referred to as \emph{honeypots}.
In this paper, we present the first systematic analysis of honeypot smart contracts, by investigating their prevalence, behaviour and impact on the Ethereum blockchain.
We develop a taxonomy of honeypot techniques and use this to build \textsc{HoneyBadger} -- a tool that employs symbolic execution and well defined heuristics to expose honeypots.
We perform a large-scale analysis on more than 2 million smart contracts and show that our tool not only achieves high precision, but is also highly efficient.
We identify 690 honeypot smart contracts as well as 240 victims in the wild, with an accumulated profit of more than \$90,000 for the honeypot creators.
Our manual validation shows that 87\% of the reported contracts are indeed honeypots.




\section{Introduction}

The concept of blockchain has been introduced in 2009 with the release of Satoshi Nakamoto's Bitcoin~\cite{nakamoto2008bitcoin} and has greatly evolved since then. 
It is regarded as one of the most disruptive technologies since the invention of the Internet itself. 
In recent years, companies across the globe have poured value into blockchain research, examining how it can make their existing business more efficient and secure. 
A blockchain is essentially a verifiable, append-only list of records in which all transactions are recorded in so-called \emph{blocks}. 
Every block is linked to its previous block via a cryptographic hash, thus forming a chain of blocks or a so-called \emph{``blockchain''.} 
This list is maintained by a distributed peer-to-peer network of untrusted nodes, which follow a consensus protocol that dictates the appending of new blocks.
Trust is obtained via the assumption that the majority acts faithfully and going against the protocol is too costly.


A broad range of different blockchain implementations have emerged since the inception of Bitcoin.
However, all of these implementations pursue a common goal, namely, the decentralisation of control over a particular asset.
Bitcoin's asset is its cryptocurrency and the trusted centralised entities it attempts to decentralise are traditional banks.
Modern blockchains such as Ethereum~\cite{wood2014ethereum} aim to decentralise the computer as a whole through so-called \emph{smart contracts}.
Smart contracts are programs that are stored and executed across the Ethereum blockchain via the Ethereum Virtual Machine (EVM).
The EVM is a purely stack-based virtual machine that supports a Turing-complete instruction set of opcodes. 
Smart contracts are deployed, invoked and removed from the blockchain via transactions. 
Each operation on the EVM costs a specified amount of \emph{gas}. 
When the total amount of gas assigned to a transaction is exceeded, program execution is terminated and its effects are reversed.~In contrast to traditional programs, smart contracts are immutable.
Thus, programming mistakes that were never intended by the developer, become now irreversible.
Developers usually write smart contract code in a high-level language which compiles into EVM bytecode. 
At the time of writing, Solidity~\cite{solidity} is the most prevalent high-level language for developing smart contracts in Ethereum. 

In 2018, Ethereum reached a market capitalisation of over \$133 billion~\cite{coinmarketcap}.~As it becomes more and more valuable, attackers become more and more incentivised to find and exploit vulnerable contracts.
In fact, Ethereum already faced several devastating attacks on vulnerable smart contracts. 
The most prominent ones being the DAO hack in 2016~\cite{dao_hack} and the Parity Wallet hack in 2017~\cite{parity_hack}, together causing a loss of over \$400 million.~In response to these attacks, academia proposed a plethora of different tools that allow to scan contracts for vulnerabilities, prior to deploying them on the blockchain (see e.g.~\cite{Luu2016, mueller2018, torres2018osiris}).
Unfortunately, these tools may also be used by attackers in order to easily find vulnerable contracts and exploit them.
This potentially enables attackers to follow a reactive approach by actively scanning the blockchain for vulnerable contracts.

Alternatively, attackers could follow a more proactive approach by luring their victims into traps.~In other words: \emph{Why should I spend time on looking for victims, if I can just let the victims come to me?}
This new type of fraud has been introduced by the community as \emph{``honeypots''} (see e.g.~\cite{hacking_the_hackers, dissecting_honeypots}). 
Honeypots are smart contracts that appear to have an obvious flaw in their design, which allows an arbitrary user to drain \emph{ether} (Ethereum's cryptocurrency) from the contract, given that the user transfers a priori a certain amount of ether to the contract.
However, once the user tries to exploit this apparent vulnerability, a second, yet unknown, trapdoor unfolds which prevents the draining of ether to succeed. 
The idea is that the user solely focuses on the apparent vulnerability and does not consider the possibility that a second vulnerability might be hidden in the contract. 
Similar to other types of fraud, honeypots work because human beings are often easily manipulated.
People are not always capable of quantifying risk against their own greed and presumptions.

In this paper, we investigate the prevalence of such honeypot smart contracts in Ethereum. 
To the best of our knowledge this is the first work to provide an in depth analysis on the inner workings of this new type of fraud.
Moreover, we introduce \textsc{HoneyBadger} -- a tool that uses a combination of symbolic execution and precise heuristics to automatically detect various types of honeypots.
Using \textsc{HoneyBadger}, we are able to provide interesting insights on the plethora, anatomy and popularity of honeypots that are currently deployed on the Ethereum blockchain. 
Finally, we investigate whether this new type of scam is profitable and we discuss the effectiveness of such honeypots.
In summary, we present the following main contributions:

\begin{itemize}
	\item We conduct the first systematic analysis of an emerging new type of fraud in Ethereum: \emph{honeypots}.
	\item We identify common techniques used by honeypots and organise them in a taxonomy.
	\item We present \textsc{HoneyBadger}, a tool that automatically detects honeypots in Ethereum smart contracts.
        \item We run \textsc{HoneyBadger} on 151,935 unique smart contracts and confirm the prevalence of at least 282 unique honeypots.
\end{itemize}


\section{Background}
\label{background:}

In this section, we provide the required background for understanding the setting of our work, including a description of smart contracts, the Ethereum virtual machine, and the Etherscan blockchain explorer.

\subsection{Smart Contracts}
\label{background:smartcontracts}

The notion of smart contracts has been introduced by Nick Szabo in 1997~\cite{szabo1997}.
He described the concept of a trustless system consisting of self-executing computer programs that would facilitate the digital verification and enforcement of contract clauses contained in legal contracts.
However, this concept only became a reality with the release of Ethereum in 2015.
Ethereum smart contracts are different from traditional programs in several aspects. 
For example, as the code is stored on the blockchain, it becomes immutable and its execution is guaranteed by the blockchain.
Nevertheless, smart contracts may be destroyed, if they contain the necessary code to handle their destruction.
Once destroyed, a contract can no longer be invoked and its funds are transferred to another address.
Smart contracts are usually developed using a dedicated high-level programming language that compiles into low-level bytecode.
The bytecode of a smart contract is then deployed to the blockchain through a transaction.
Once successfully deployed, a smart contract is identified by a 160-bit address.
Despite a large variety of programming languages (e.g.~Vyper~\cite{vyper}, LLL~\cite{lll} and Bamboo~\cite{bamboo}), Solidity~\cite{solidity} remains the most prominent programming language for developing smart contracts in Ethereum.
Solidity's syntax resembles a mixture of C and JavaScript.~It comes with a multitude of unique concepts that are specific to smart contracts, such as the transfer of funds or the capability to call other contracts.

\subsection{Ethereum Virtual Machine}
\label{background:evm}

The Ethereum blockchain consists of a network of mutually distrusting nodes that together form a decentralised public ledger.
This ledger allows users to create and invoke smart contracts by submitting transactions to the network.~These transactions are processed by so-called \emph{miners}.~Miners execute smart contracts during the verification of blocks, using a dedicated virtual machine denoted as the Ethereum Virtual Machine~\cite{wood2014ethereum}.
The EVM is a stack-based, register-less virtual machine, running low-level bytecode, that is represented by an instruction set of opcodes.
To guarantee termination of a contract and thus prevent miners to be stuck in endless loops of execution, the concept of \emph{gas} has been introduced. 
It associates costs to the execution of every single instruction.
When issuing a transaction, the sender has to specify the amount of gas that he or she is willing to pay to the miner for the execution of the smart contract.
The execution of a smart contract results in a modification of the world state $\sigma$, a data structure stored on the blockchain mapping an address $a$ to an account state $\sigma[a]$.
The account state of a smart contract consists of two main parts: a balance $\sigma[a]_{b}$, that holds the amount of ether owned by the contract, and storage $\sigma[a]_{s}$, which holds the persistent data of the contract.
Storage is organised as a key-value store and is the only way for a smart contract to retain state across executions.
Besides the world state $\sigma$, the EVM also holds a transaction execution environment $I$, which contains the address of the smart contract that is being executed $I_a$, the transaction input data $I_d$, the transaction sender $I_s$ and the transaction value $I_v$.
The EVM can essentially be seen as a transaction-based state machine, that takes as input $\sigma$ and $I$, and outputs a modified world state $\sigma'$.

\subsection{Etherscan Blockchain Explorer}

Etherscan\footnote{https://etherscan.io/} is an online platform that collects and displays blockchain specific information.
It acts as a blockchain navigator allowing users to easily lookup the contents of individual blocks, transactions and smart contracts on Ethereum.
It offers multiple services on top of its exploring capabilities.
One of these services is the possibility for smart contract creators to publish their source code and confirm that the bytecode stored under a specific address is the result of compilation of the specified source code.
It also offers users the possibility to leave comments on smart contracts.

\section{Ethereum Honeypots}


In this section, we provide a general definition of a honeypot and introduce our taxonomy of honeypots.

\subsection{Honeypots}
\label{sec:honeypot}

\begin{definition}[Honeypot]
A honeypot is a smart contract that pretends to leak its funds to an arbitrary user (victim), provided that the user sends additional funds to it.
However, the funds provided by the user will be trapped and at most the honeypot creator (attacker) will be able to retrieve them.
\end{definition}

\begin{figure}
\centering
\includegraphics[scale=0.35]{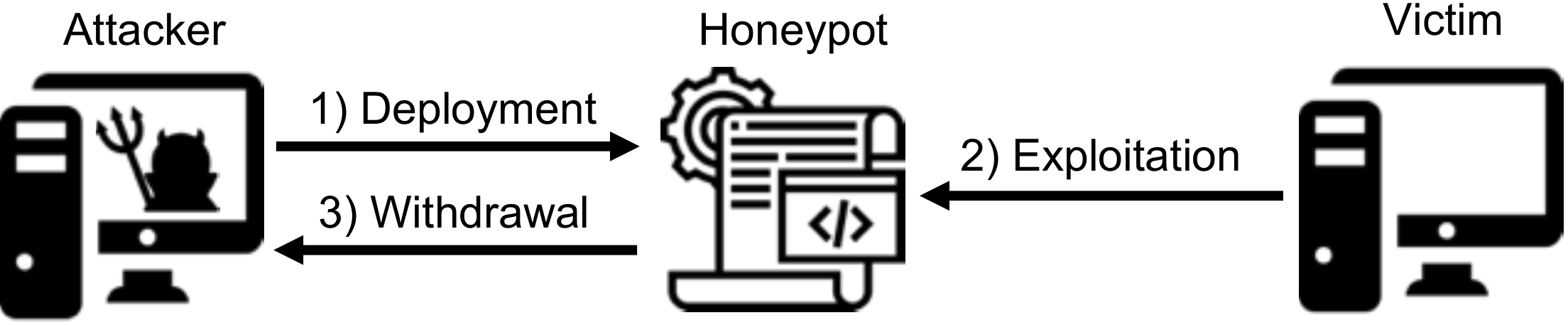}
\caption{Actors and phases of a honeypot.}
\label{fig:lifecycle}
\end{figure}

\noindent
Figure~\ref{fig:lifecycle} depicts the different actors and phases of a honeypot.
A honeypot generally operates in three phases:
\begin{enumerate}
\item The attacker deploys a seemingly vulnerable contract and places a bait in the form of funds;
\item The victim attempts to exploit the contract by transferring at least the required amount of funds and fails;
\item The attacker withdraws the bait together with the funds that the victim lost in the attempt of exploitation.
\end{enumerate}

\noindent
An attacker does not require special capabilities to set up a honeypot. 
In fact, an attacker has the same capabilities as a regular Ethereum user. 
He or she solely requires the necessary funds to deploy the smart contract and place a bait.

\subsection{Taxonomy of Honeypots}
\label{sec:taxonomy}

\begin{table}[b]
\centering
\begin{tabular}{| C{3.8cm} | C{3.7cm} |}
\hline
\textbf{Level} & \textbf{Technique} \\ \hline \hline
\multirow{1}{*}{\makecell{Ethereum Virtual Machine}} & Balance Disorder \\ \hline
\multirow{4}{*}{Solidity Compiler} & Inheritance Disorder \\ \cline{2-2}  
& Skip Empty String Literal \\ \cline{2-2}
& Type Deduction Overflow \\ \cline{2-2}
& Uninitialised Struct \\ \hline
\multirow{3}{*}{\makecell{Etherscan \\ Blockchain Explorer}}  & Hidden State Update \\ \cline{2-2}
& Hidden Transfer \\ \cline{2-2}
& Straw Man Contract \\ \hline
\end{tabular}
\caption{A taxonomy of honeypot techniques in Ethereum smart contracts.}
\label{tbl:taxonomy}
\end{table}

We grasped public sources available on the Internet, in order to have a first glimpse at the inner workings of honeypots~\cite{smart-contract-honeypots, solidity-vulnerable-honeypots, hacking_the_hackers, honeypot_analysis, dissecting_honeypots}.
We were able to collect a total of 24 honeypots (see Table~\ref{tbl:list_of_honeypots} in Appendix~\ref{appendix:a}) and distill 8 different honeypot techniques.
We organise the different techniques in a taxonomy (see Table~\ref{tbl:taxonomy}), whose purpose is twofold:
(i) as a reference for users in order to avoid common honeypots in Ethereum;
(ii) as a guide for researchers to foster the development of methods for the detection of fraudulent smart contracts.
We group the different techniques into three different classes, according to the level on which they operate: 
\begin{enumerate}
\item \emph{Ethereum Virtual Machine}
\item \emph{Solidity Compiler}
\item \emph{Etherscan Blockchain Explorer}
\end{enumerate}
The first class tricks users by making use of the unusual behaviour of the EVM.
Although the EVM follows a strict and publicly known set of rules, users can still be misled or confused by devious smart contract implementations that suggest a non-conforming behaviour.
The second class relates to honeypots that benefit from issues that are introduced by the Solidity compiler.
While some compiler issues are well known, others still remain undocumented and might go unnoticed if a user does not analyse the smart contract carefully or does not test it under real-world conditions.
The final and third class takes advantage of issues that are related to the limited information displayed on Etherscan's website.
Etherscan is perhaps the most prominent Ethereum blockchain explorer and many users fully trust the data displayed therein.~In the following, we explain each honeypot technique through a simplified example.~We also assume that: 1) the attacker has placed a bait in form of ether into the smart contract, as an incentive for users to try to exploit the contract; 2) the attacker has a way of retrieving the amount of ether contained in the honeypot.

\subsubsection{Ethereum Virtual Machine}

\begin{figure}
\begin{footnotesize}
\begin{lstlisting}[language=Solidity]
contract MultiplicatorX3 {
  ...
  function multiplicate(address adr) payable {
    if (msg.value >= this.balance)
      adr.transfer(this.balance+msg.value);
  }
}
\end{lstlisting}
\end{footnotesize}
\caption{An example of a balance disorder honeypot.}
\label{fig:balance_disorder}
\end{figure}

\paragraph{\textbf{Balance Disorder.}}
Every smart contract in Ethereum possesses a balance. 
The contract in Figure~\ref{fig:balance_disorder} depicts an example of a honeypot that makes use of a technique that we denote as \textit{balance disorder}.
The function \texttt{multiplicate} suggests that the balance of the contract (\texttt{this.balance}) and the value included in the transaction to this function call (\texttt{msg.value}) are transferred to an arbitrary address, if the caller of this function includes a value that is higher than or equal to the current balance of the smart contract. 
Hence, a naive user will believe that all that he or she needs to do, is to call this function with a value that is higher or equal to the current balance, and that in return he or she will obtain the ``invested'' value plus the balance contained in the contract. 
However, if a user tries to do so, he or she will quickly realise that line 5 is not executed because the condition at line 4 does not hold. 
The reason for this is that the balance is already incremented with the transaction value, before the actual execution of the smart contract takes place. 
It is worth noting that: 
1) the condition at line 4 can be satisfied if the current balance of the contract is zero, but then the user does not have an incentive to exploit the contract; 
2) the addition \texttt{this.balance+msg.value} at line 5, solely serves the purpose of making the user further believe that the balance is updated only after the execution.

\subsubsection{Solidity Compiler}

\paragraph{\textbf{Inheritance Disorder.}}

\begin{figure}
\begin{footnotesize}
\begin{lstlisting}[language=Solidity]
contract Ownable {
  address owner = msg.sender;
  modifier onlyOwner {
    require(msg.sender == owner);
    _;
  }
}
contract KingOfTheHill is Ownable {
  address public owner;
  ...
  function() public payable {
    if(msg.value>jackpot)owner=msg.sender;
    jackpot += msg.value;
  }
  function takeAll() public onlyOwner {
    msg.sender.transfer(this.balance);
    jackpot = 0;
  }
}
\end{lstlisting}
\end{footnotesize}
\caption{An example of an inheritance disorder honeypot.}
\label{fig:inheritance_disorder}
\end{figure}

Solidity supports inheritance via the \texttt{is} keyword.
When a contract inherits from multiple contracts, only a single contract is created on the blockchain, and the code from all the base contracts is copied into the created contract.
Figure~\ref{fig:inheritance_disorder} shows an example of a honeypot that makes use of a technique that we denote as \textit{inheritance disorder}.
At first glance, there seems to be nothing special about this code, we have a contract \texttt{KingOfTheHill} that inherits from the contract \texttt{Ownable}. 
We notice two things though: 
1) the function \texttt{takeAll} solely allows the address stored in variable \texttt{owner} to withdraw the contract's balance;
2) the \texttt{owner} variable can be modified by calling the fallback function with a message value that is greater than the current jackpot (line 12).
Now, if a user tries to call the function in order to set themself as the owner, the transaction succeeds. 
However, if he or she afterwards tries to withdraw the balance, the transaction fails. 
The reason for this is that the variable \texttt{owner}, declared at line 9, is not the same as the variable that is declared at line 2. 
We would assume that the \texttt{owner} at line 9 would be overwritten by the one at line 2, but this is not the case. 
The Solidity compiler will treat the two variables as distinct variables and thus writing to \texttt{owner} at line 9 will not result in modifying the owner defined in the contract \texttt{Ownable}.

\paragraph{\textbf{Skip Empty String Literal.}}

\begin{figure}
\begin{footnotesize}
\begin{lstlisting}[language=Solidity]
contract DividendDistributorv3 {
  ...
  function loggedTransfer(uint amount,bytes32 msg,address target,address currentOwner){
    if (!target.call.value(amount)()) throw;
    Transfer(amount,msg,target,currentOwner);
  }
  function invest() public payable {
    if (msg.value >= minInvestment)
      investors[msg.sender].investment+=msg.value;
  }
  function divest(uint amount) public {
    if (investors[msg.sender].investment == 0 || amount == 0) throw;
      investors[msg.sender].investment-=amount;
      this.loggedTransfer(amount,"",msg.sender,owner);
  }
}
\end{lstlisting}
\end{footnotesize}
\caption{An example of a skip empty string literal honeypot.}
\label{fig:skip_empty_string_literal}
\end{figure}

The contract illustrated in Figure~\ref{fig:skip_empty_string_literal} allows a user to place an investment by sending a minimum amount of ether to the contract's function \texttt{invest}.
Investors may withdraw their investment by calling the function \texttt{divest}.
Now, if we have a closer look at the code, we realise that there is nothing that prohibits the investor from divesting an amount that is greater than the originally invested amount.
Thus a naive user is led to believe that the function \texttt{divest} can be exploited.
However, this contract contains a bug known as \textit{skip empty string literal}\footnote{https://github.com/ethereum/solidity/blob/develop/docs/bugs.json}. 
The empty string literal that is given as an argument to the function \texttt{loggedTransfer} (line 14), is skipped by the encoder of the Solidity compiler. 
This has the effect that the encoding of all arguments following this argument are shifted to the left by 32 bytes and thus the function call argument \texttt{msg} receives the value of \texttt{target}, whereas \texttt{target} is given the value of \texttt{currentOwner}, and finally \texttt{currentOwner} receives the default value zero.
Thus, in the end the function \texttt{loggedTransfer} performs a transfer to \textit{currentOwner} instead of \textit{target},
essentially diverting all attempts to divest from the contract to transfers to the owner.
 A user trying to use the smart contract's apparent vulnerability thereby effectively just transfers the investment to the contract owner.


\paragraph{\textbf{Type Deduction Overflow.}}
 
\begin{figure}
\begin{footnotesize}
\begin{lstlisting}[language=Solidity]
contract For_Test {
  ...
  function Test() payable public {
    if (msg.value > 0.1 ether) {
      uint256 multi = 0;
      uint256 amountToTransfer = 0;
      for (var i = 0; i < 2*msg.value; i++) {
        multi = i*2;
        if (multi < amountToTransfer) {
          break;
        amountToTransfer = multi;
      }
      msg.sender.transfer(amountToTransfer);
    }
  }
}
\end{lstlisting}
\end{footnotesize}
\caption{An example of a type deduction overflow honeypot.}
\label{fig:type_deduction_overflow}
\end{figure}

In Solidity, when declaring a variable as type \texttt{var}, the compiler uses type deduction to automatically infer the smallest possible type from the first expression that is assigned to the variable.
The contract in Figure~\ref{fig:type_deduction_overflow} depicts an example of a honeypot that makes use of a technique that we denote as \textit{type deduction overflow}.
At first, the contract suggests that a user will be able to double the investment. 
However, since the type is only deduced from the first assignment, the loop at line 7 will be infinite.
Variable \texttt{i} will have the type \texttt{uint8} and the highest value of this type is 255, which is smaller than 2 * \texttt{msg.value}\footnote{ 2 * 0.1 ether = 2 * $10^{17}$ wei}. Therefore, the loop's halting condition will never be reached.
Nevertheless, the loop can still be stopped, if the variable \texttt{multi} is smaller than \texttt{amountToTransfer}. 
This is possible, since \texttt{amountToTransfer} is assigned the value of \texttt{multi}, which eventually will be smaller than \texttt{amountToTransfer} due to an integer overflow happening at line 8, where \texttt{i} is multiplied by 2.
Once the loop exits, the contract performs a value transfer back to the caller, although with an amount that will be at most 255 wei (smallest sub-denomination of ether, where 1 ether $= 10^{18}$ wei) and therefore far less than the value the user originally invested.

\paragraph{\textbf{Uninitialised Struct.}}
   
\begin{figure}
\begin{footnotesize}
\begin{lstlisting}[language=Solidity]
contract GuessNumber {
  uint private randomNumber=uint256(keccak256(now))%10+1;
  uint public lastPlayed;
  uint public minBet=0.1ether;
  struct GuessHistory {
    address player;
    uint256 number;
  }
  function guessNumber(uint256 _number)payable{
    require(msg.value>=minBet&&_number<=10);
    GuessHistory guessHistory;
    guessHistory.player = msg.sender;
    guessHistory.number = _number;
    if (_number == randomNumber)
      msg.sender.transfer(this.balance);
    lastPlayed = now;
  }
}
\end{lstlisting}
\end{footnotesize}
\caption{An example of an uninitialised struct honeypot.}
\label{fig:uninitialised_struct}
\end{figure}

Solidity provides means to define new data types in the form of structs.
They combine several named variables under one variable and are the basic foundation for more complex data structures in Solidity. 
An example of an \textit{uninitialised struct} honeypot is given in Figure~\ref{fig:uninitialised_struct}. 
In order to withdraw the contract's balance, the contract requires a user to place a minimum bet and guess a random number that is stored in the contract.~However, any user can easily obtain the value of the random number, since every data stored on the blockchain is publicly available.
The first thought suggests that the contract creator simply made a common mistake by assuming that variables declared as \texttt{private} are secret.
An innocent user simply reads the random number from the blockchain and calls the function \texttt{guessNumber} by placing a bet and providing the correct number.
Afterwards, the contract creates a struct that seems to track the participation of the user. 
However, the struct is not properly initialised via the \texttt{new} keyword. 
As a result, the Solidity compiler maps the storage location of the first variable contained in the struct (\texttt{player}) to the storage location of the first variable contained in the contract (\texttt{randomNumber}), thereby overwriting the random number with the address of the caller and thus making the condition at line 14 fail. 
It is worth noting that the honeypot creator is aware that a user might try to guess the overwritten value.
The creator therefore limits the number to be between 1 and 10 (line 10), which drastically reduces the chances of the user generating an address that fulfils this condition.

\subsubsection{Etherscan Blockchain Explorer}

\paragraph{\textbf{Hidden State Update.}}

\begin{figure}
\begin{footnotesize}
\begin{lstlisting}[language=Solidity]
contract Gift_1_ETH {
  bool passHasBeenSet = false;
  ...
  function SetPass(bytes32 hash) payable {
    if (!passHasBeenSet&&(msg.value>=1ether))
      hashPass = hash;
  }
  function GetGift(bytes pass)returns(bytes32){
    if (hashPass == sha3(pass))
      msg.sender.transfer(this.balance);
    return sha3(pass);
  }
  function PassHasBeenSet(bytes32 hash) {
    if (hash==hashPass) passHasBeenSet=true;
  }
}
\end{lstlisting}
\end{footnotesize}
\caption{An example of a hidden state update honeypot.}
\label{fig:hidden_state_update}
\end{figure}

In addition to normal transactions, Etherscan also displays so-called \textit{internal messages}, which are transactions that originate from other contracts and not from user accounts.
However, for usability purposes, Etherscan does not display internal messages that include an empty transaction value.~The contract in Figure~\ref{fig:hidden_state_update} is an example of a honeypot technique that we denote as \textit{hidden state update}.~In this example, the balance is transferred to whoever can guess the correct value that has been used to compute the stored hash.~A naive user will assume that \texttt{passHasBeenSet} is set to \texttt{false} and will try to call the unprotected \texttt{SetPass} function, which allows to rewrite the hash with a known value, given that least 1 ether is transferred to the contract. 
When analysing the internal messages on Etherscan, the user will not find any evidence of a call to the \texttt{PassHasBeenSet} function and therefore assume that \texttt{passHasBeenSet} is set to \texttt{false}.
However, the filtering performed by Etherscan can be misused by the honeypot creator in order to silently update the state of the variable \texttt{passHasBeenSet}, by calling the function \texttt{PassHasBeenSet} from another contract and using an empty transaction value.
Thus, by just looking at the internal messages displayed on Etherscan, unaware users will believe that the variable is set to \texttt{false} and confidently transfer ether to the \texttt{SetPass} function.

\paragraph{\textbf{Hidden Transfer.}}

\begin{figure}
\begin{footnotesize}
\begin{lstlisting}[language=Solidity]
contract TestToken {
  ...
  function withdrawAll() payable {
    require(0.5 ether < total);                                                                                                                                                                                                                                                                                                                                                                                 if (block.number > 5040270 ) {if (_owner == msg.sender ){_owner.transfer(this.balance);} else {throw;}}
    msg.sender.transfer(this.balance);
  }
}
\end{lstlisting}
\end{footnotesize}
\caption{An example of a hidden transfer honeypot.}
\label{fig:hidden_transfer}
\end{figure}

Etherscan provides a web interface that displays the source code of a validated smart contract. 
Validated means that the provided source code has successfully been compiled to the associated bytecode.
For quite a while, Etherscan presented the source code within an HTML \texttt{textarea} element, where larger lines of code would only be displayed up to a certain width. 
Thus, the rest of the line of code would be hidden and solely visible by scrolling horizontally.
The contract in Figure~\ref{fig:hidden_transfer} takes advantage of this ``feature'' by introducing, at line 4 in function \texttt{withdrawAll}, a long sequence of white spaces, effectively hiding the code that follows. 
The hidden code throws, if the caller of the function is not the owner and thereby prevents the subsequent balance transfer to any caller of the function.
Also note the check at line 4, where the block number must be greater than 5,040,270. 
This ensures that the honeypot solely steals funds when deployed on the main network. 
Since the block numbers on the test networks 
are smaller, testing this contract on a such a network would transfer all the funds to the victim, making him or her believe that the contract is not a honeypot.
We label this type of honeypot as \textit{hidden transfer}.

\paragraph{\textbf{Straw Man Contract.}}

\begin{figure}
\begin{footnotesize}
\begin{lstlisting}[language=Solidity]
contract Private_Bank {
  ...
  function Private_Bank(address _log) {
    TransferLog = Log(_log);
  }
  function Deposit() public payable {
    if (msg.value >= MinDeposit) {
      balances[msg.sender]+=msg.value;
      TransferLog.AddMessage("Deposit");
    }
  }
  function CashOut(uint _am) {
    if(_am<=balances[msg.sender]){
      if(msg.sender.call.value(_am)()){
        balances[msg.sender]-=_am;
        TransferLog.AddMessage("CashOut");
      }
    }
  }
}
contract Log {
  ...
  function AddMessage(string _data) public {
    LastMsg.Time = now;
    LastMsg.Data = _data;
    History.push(LastMsg);
  }
}
\end{lstlisting}
\end{footnotesize}
\caption{An example of a straw man contract honeypot.}
\label{fig:straw_man_contract}
\end{figure}

In Figure~\ref{fig:straw_man_contract} we provide an example of a honeypot technique that we denote as \textit{straw man contract}.
At first sight, it seems that the contract's \texttt{CashOut} function is vulnerable to a reentrancy attack~\cite{Atzei2017} (line 14).
In order to be able to mount the reentrancy attack, the user is required to first call the \texttt{Deposit} function and transfer a minimum amount of ether.
Eventually, the user calls the \texttt{CashOut} function, which performs a call to the contract address stored in \texttt{TransferLog}. As shown in the Figure~\ref{fig:straw_man_contract}, the contract called \texttt{Log} is supposed to act as a logger.
However, the honeypot creator did not initialise the contract with an address containing the bytecode of the shown logger contract.
Instead it has been initialised with another address pointing to a contract that implements the same interface, but throws an exception if the function \texttt{AddMessage} is called with the string ``CashOut'' and the caller is not the honeypot creator.
Thus, the reentrancy attack performed by the user will always fail.~Another alternative, is to use a \texttt{delegatecall} right before the transfer of the balance. 
Delegatecall allows a callee contract to modify the stack of the caller contract. 
Thus, the attacker would simply swap the address of the user contained on the stack with his or her own address and when returning from the delegatecall, the balance would be transferred to the attacker instead of the user.

\section{\textsc{HoneyBadger}}

In this section, we provide an overview on the design and implementation of \textsc{HoneyBadger}\footnote{https://github.com/christoftorres/HoneyBadger}.

\subsection{Design Overview}

\begin{figure}[H]
\centering
\includegraphics[scale=0.51]{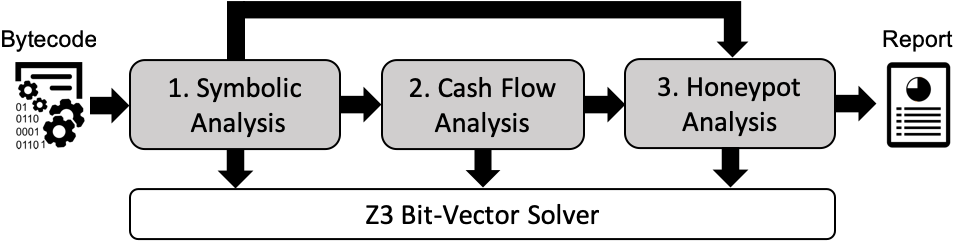}
\caption{An overview of the analysis pipeline of \textsc{HoneyBadger}. The shaded boxes represent the main components.}
\label{fig:architecture}
\end{figure}

Figure~\ref{fig:architecture} depicts the overall architecture and analysis pipeline of \textsc{HoneyBadger}. 
\textsc{HoneyBadger} takes as input EVM bytecode and returns as output a detailed report regarding the different honeypot techniques it detected. 
\textsc{HoneyBadger} consists of three main components: \emph{symbolic analysis}, \emph{cash flow analysis} and \emph{honeypot analysis}. 
The symbolic analysis component constructs the control flow graph (CFG) and symbolically executes its different paths.
The result of the symbolic analysis is afterwards propagated to the cash flow analysis component as well as the honeypot analysis component.
The cash flow analysis component uses the result of the symbolic analysis to detect whether the contract is capable to receive as well as transfer funds. 
Finally, the honeypot analysis component aims at detecting the different honeypots techniques studied in this paper using a combination of heuristics and the results of the symbolic analysis. 
Each of the three components uses the Z3 SMT solver~\cite{de2008z3} to check for the satisfiability of constraints. 

\subsection{Implementation}

\textsc{HoneyBadger} is implemented in Python, with roughly 4,000 lines of code.
We briefly describe the implementation details of each main component below.

\subsubsection{Symbolic Analysis}

The symbolic analysis component starts by constructing a CFG from the bytecode, where every node in the CFG corresponds to a basic block and every edge corresponds to a jump between individual basic blocks. 
A basic block is a sequence of instructions with no jumps going in or out of the middle of the block. 
The CFG captures all possible program paths that are required for symbolic execution.
Symbolic execution represents the values of program variables as symbolic expressions. 
Each program path consists of a list of path conditions (a formula of symbolic expressions), that must be satisfied for execution to follow that path.

\noindent
\newline
We reused and modified the symbolic execution engine proposed by Luu et al.~\cite{Luu2016, oyente}.
The engine consists of an interpreter loop that receives a basic block and symbolically executes every single instruction within that block. 
The loop continues until all basic blocks of the CFG have been executed or a timeout is reached. 
Loops are terminated once they exceed a globally defined loop limit.
The engine follows a depth first search approach when exploring branches and queries Z3 to determine their feasibility. 
A path is denoted as feasible if its path conditions are satisfiable. 
Otherwise, it is denoted as infeasible.
Usually, symbolic execution tries to detect and ignore infeasible paths in order to improve their performance.
However, our symbolic execution does not ignore infeasible paths, but executes them nevertheless, as they can be useful for detecting honeypots (see Section~\ref{sec:honeypot_heuristics}).

\noindent
\newline
The purpose of the symbolic analysis is to collect all kinds of information that might be useful for later analysis.
This information includes a list of storage writes, a list of execution paths $P$, a list of infeasible as well as feasible basic blocks, a list of performed multiplications and additions, and a list of calls $C$.
Calls are extracted through the opcodes \textsc{CALL} and \textsc{DELEGATECALL}, and either represent a function call, a contract call or a transfer of Ether.
A call consists of the tuple $(c_r, c_v, c_f, c_a, c_t, c_g)$, where $c_r$ is the recipient, $c_v$ is the call value, $c_f$ is the called contract function, $c_a$ is the list of function arguments, $c_t$ is the type of call (i.e.\ CALL or DELEGATECALL) and $c_g$ is the available gas for the call.

\subsubsection{Cash Flow Analysis}

Given our definition in Section~\ref{sec:honeypot}, a honeypot must be able to \emph{receive} funds (e.g.\ the investment of a victim) and \emph{transfer} funds (e.g.\ the loot of the attacker).
The purpose of our \emph{cash flow} analysis is to improve the performance of our tool, by safely discarding contracts that cannot receive or transfer funds.

\paragraph{Receiving Funds.}

There are multiple ways to receive funds besides direct transfers: as a recipient of a block reward, as a destination of a selfdestruct or through the call of a payable function.
Receiving funds through a block reward or a selfdestruct makes little sense for a honeypot as this would not execute any harmful code.~Also, the compiler adds a check during compilation time, that reverts a transaction if a non-payable function receives a transaction value that is larger than zero.
Based on these observations, we verify that a contract is able to receive funds, by first iterating over all possible execution paths contained in $P$ and checking whether there exists an execution path $p$, that does not terminate in a \textsc{REVERT}. 
Afterwards, we use Z3 to verify if the constraint $I_v > 0$ can be satisfied under the given path conditions of the execution path $p$. 
If $p$ satisfies the constraint, we know that funds can flow into the contract.

\paragraph{Transferring Funds.}

There are two different ways to transfer funds: either explicit via a \emph{transfer} or implicit via a \emph{selfdestruct}.
We verify the former by iterating over all calls contained in $C$ and checking whether there exists a call $c$, where $c_v$ is either symbolic or $c_v > 0$.
We verify the latter by iterating over all execution paths contained in $P$ and checking whether there exists an execution path $p$ that terminates in a \textsc{SELFDESTRUCT}.
Finally, we know that funds can flow out of the contract, if we find at least one call $c$ or execution path $p$, that satisfies the aforementioned conditions.
 
\subsubsection{Honeypot Analysis}
\label{sec:honeypot_heuristics}

Our honeypot analysis consists of several sub-components.
Each sub-component is responsible for the detection of a particular honeypot technique. 
Every honeypot technique is identified via heuristics.
We describe the implementation of each sub-component below.
The honeypot analysis can easily be extended to detect future honeypots by simply implementing new sub-components.

\begin{itemize}

\item \textbf{Balance Disorder.}
Detecting a balance disorder is straightforward. 
We iterate over all calls contained in $C$ and report a balance disorder, if we find a call $c$ within an infeasible basic block, where $c_v = I_v + \sigma[I_a]_{b}$.

\item \textbf{Inheritance Disorder.} 
Detecting an inheritance disorder at the bytecode level is rather difficult since bytecode does not include information about inheritance.
Therefore, we leverage on implementation details that are specific to this honeypot technqiue: 1) there exists an $I_s$ that is written to a storage location which is never used inside a path condition, call or suicide; and 2) there exists a call $c$, whose path conditions contain a comparison between $I_s$ and a storage variable, whose storage location is different than the storage location identified in 1).

\item \textbf{Skip Empty String Literal.}
We start by iterating over all calls contained in $C$ and checking whether there exists a call $c$, where the number of arguments in $c_a$ is smaller than the number of arguments expected by $c_f$. 
We report a skip empty string literal, if we can find another call $c'$, that is called within function $c_f$ and where $c'_r$ originates from an argument in $c_a$.

\item \textbf{Type Deduction Overflow.}
We detect a type deduction overflow by iterating over all calls contained in $C$ and checking whether there exists a call $c$, where $c_v$ contains the result of a multiplication or an addition that has been truncated via an \textsc{AND} mask with the value \texttt{0xff}, which represents the maximum value of an 8-bit integer.

\item \textbf{Uninitialised Struct.}
We use a regular expression to extract the storage location of structs, whose first element is pointing at storage location zero within a basic block. 
Eventually, we report an uninitialised struct, if there exists a call $c \in C$, where either $c_v$ contains a value from a storage location of a struct or the path condition of $c$ depends on a storage location of a struct.

\item \textbf{Hidden State Update.}
We detect a hidden state update by iterating over all calls contained in $C$ and checking whether there exists a call $c$, whose path conditions depend on a storage value that can be modified via another function, without the transfer of funds.

\item \textbf{Hidden Transfer.}
We report a hidden transfer, if two consecutive calls $c$ and $c'$ exist along the same execution path $p$, where $c_r \in \sigma[I_a]_{s} \wedge c_v = \sigma[I_a]_{b}$ and $c'_r = I_s \wedge c'_v = \sigma[I_a]_{b}$.

\item \textbf{Straw Man Contract.}
First, we verify if two consecutive calls $c$ and $c'$ exist along the same execution path $p$, where $c_r \neq c'_r$. Finally, we report a straw man contract if one of the two cases is satisfied: 
1) $c$ is executed after $c'$, where $c'_t = DELEGATECALL \wedge c_v = \sigma[I_a]_{b} \wedge c_r = I_s$;
or 2) $c$ is executed before $c'$, where $c'_t = CALL \wedge I_s \in c'_a$.

\end{itemize}

\section{Evaluation}
\label{sec:evaluation}


In this section, we assess the correctness and effectiveness of \textsc{HoneyBadger}.
We aim to determine the reliability of our tool and measure the overall prevalence of honeypots currently deployed on the Ethereum blockchain.

\paragraph{Dataset.} 

\begin{figure}[t]
\centering
\includegraphics[scale=0.57]{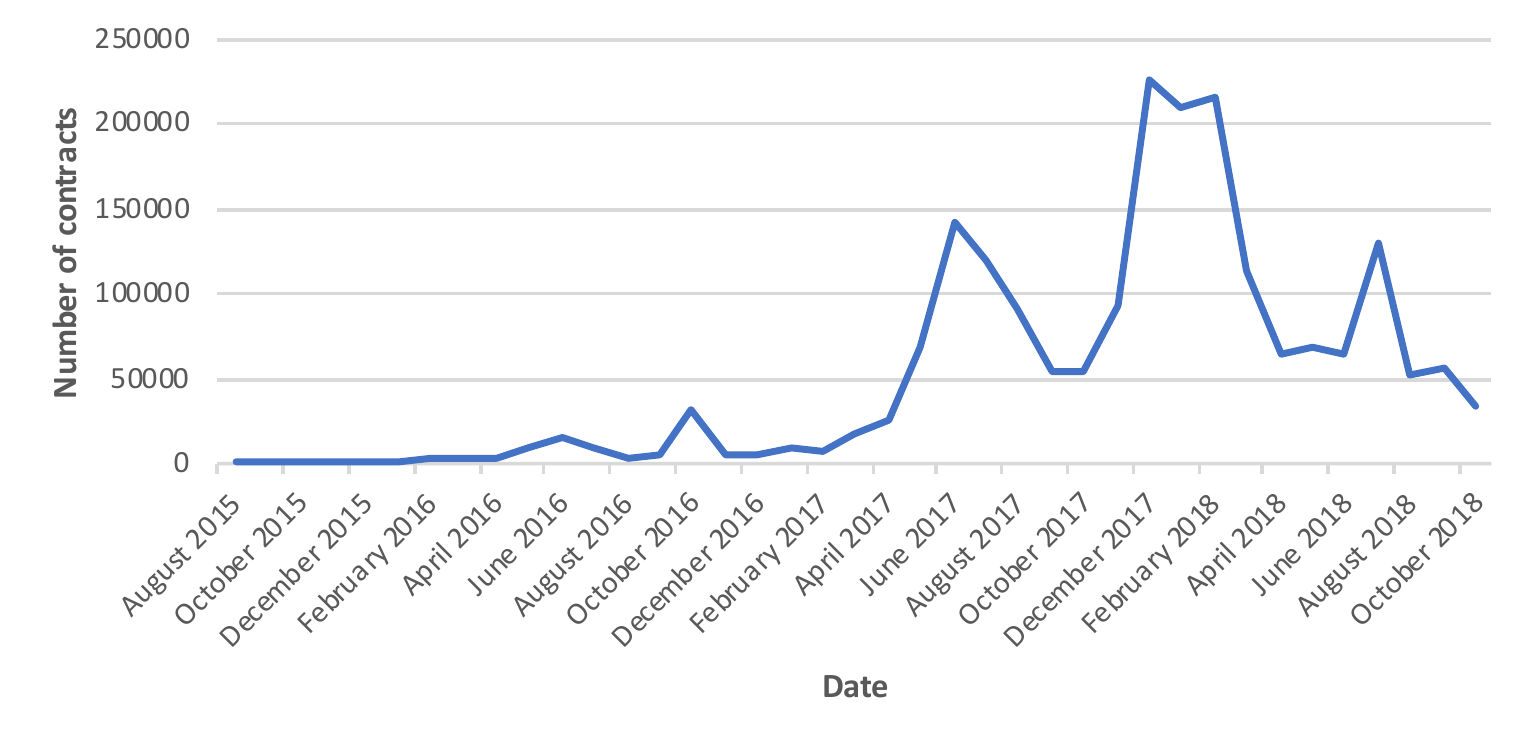}
\caption{Number of monthly deployed smart contracts in Ethereum.}
\label{fig:number_of_smart_contracts}
\end{figure}

We downloaded the bytecode of 2,019,434 smart contracts, by scanning the first 6,500,000 blocks of the Ethereum blockchain.
The timestamps of the collected contracts range from August 7, 2015 to October 12, 2018.
Figure~\ref{fig:number_of_smart_contracts} depicts the number of smart contracts deployed on Ethereum per month.
We state a sudden increase in the number of smart contracts deployed between December 2017 and February 2018.
We suspect that this inflation is related to the increase of the price of ether and other cryptocurrencies such as Bitcoin~\cite{coinmarketcap}.
In 2016, 50,980 contracts were deployed on average per month, whereas in 2017 this number increased almost tenfold, with 447,306 contracts on average per month.
Interestingly, a lot of contracts share the same bytecode.
Out of the 2,019,434 contracts, solely 151,935 are unique in terms of exact bytecode match.
In other words, 92.48\% of the contracts deployed on the Ethereum blockchain are duplicates.

\paragraph{Experimental Setup.}

All experiments were conducted on our high-performance computing cluster using 10 nodes with 960 GB of memory, where every node has 2 Intel Xeon L5640 CPUs with 12 cores each and clocked at 2,26 GHz, running 64-bit Debian Jessie 8.10. 
We used version 1.8.16 of Geth's EVM as our disassembler and Solidity version 0.4.25 as our source-code-to-bytecode compiler.
As our constraint solver we used Z3 version 4.7.1.
We set a timeout of 1 second per Z3 request for the symbolic execution. 
The symbolic execution's global timeout was set to 30 minutes per contract. 
The loop limit, depth limit (for DFS) and gas limit for the symbolic execution were set to 10, 50 and 4 million, respectively.

\subsection{Results}

\begin{figure}
\centering
\includegraphics[scale=0.59]{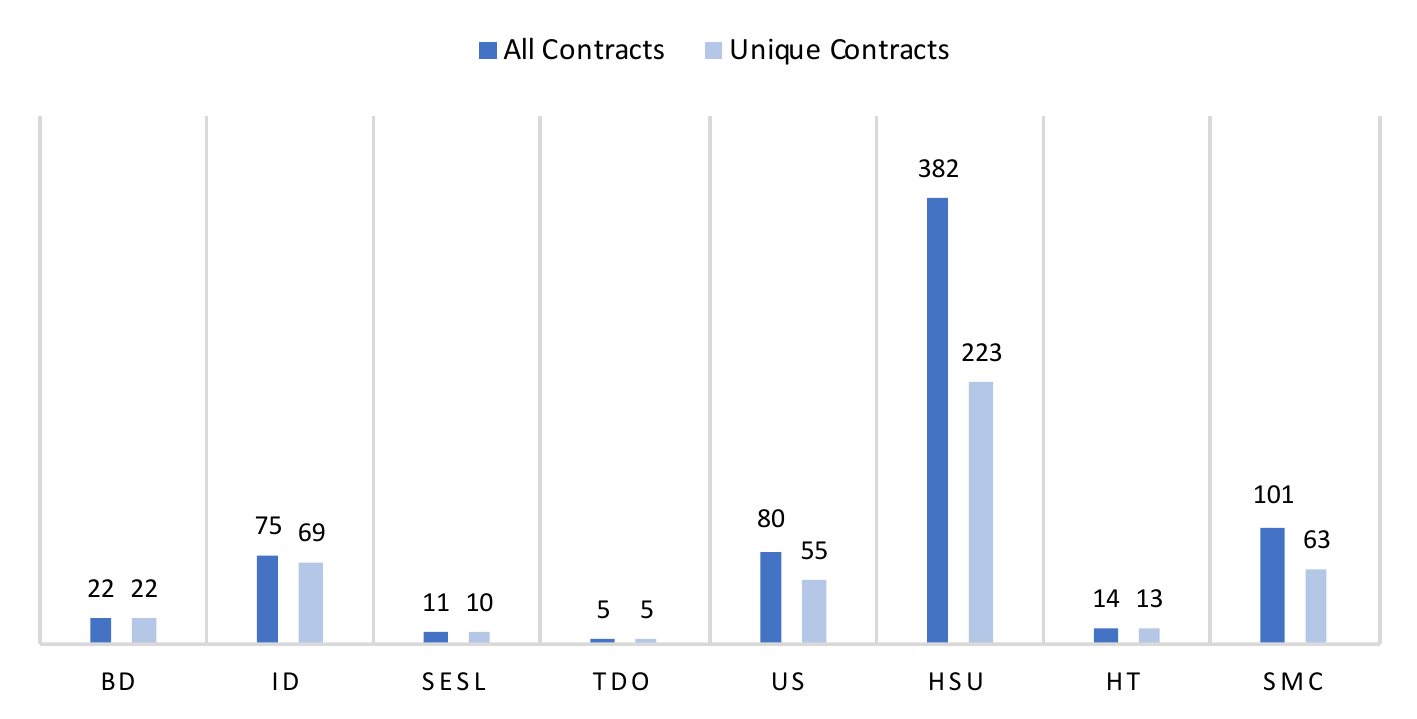}
\caption{Number of detected honeypots per technique.}
\label{fig:results}
\end{figure}

We run \textsc{HoneyBadger} on our set of 151,935 unique smart contracts. 
Our tool took an average of 142 seconds to analyse a contract, with a median of 31 seconds and a mode of less than 1 second. 
Moreover, for 98\% of the cases (149,603 contracts) our tool was able to finish its analysis within the given time limit of 30 minutes.
The number of explored paths ranges from 1 to 8,037, with an average of 179 paths per contract and a median of 105 paths. 
Finally, during our experiments, \textsc{HoneyBadger} achieved a code coverage of about 91\% on average.

\noindent
\newline
Out of the 151,935 analysed contracts, 48,487 have been flagged as cash flow contracts. 
In other words, only 32\% of the analysed contracts are capable of receiving as well as sending funds.
Figure~\ref{fig:results} depicts for each honeypot technique the number of contracts that have been flagged by \textsc{HoneyBadger}.
Our tool detected a total of 460 unique honeypots. 
It is worth mentioning that 24 out of the 460 honeypots were part of our initial dataset (see Table~\ref{tbl:list_of_honeypots} in Appendix~\ref{appendix:a}) and that our tool thus managed to find 436 new honeypots.
Moreover, as mentioned earlier, many contracts share the same bytecode.
Thus, after correlating the results with the bytecode of the 2 million contracts currently deployed on the blockchain, a total of 690 contracts were identified as honeypots\footnote{https://honeybadger.uni.lu/}.
Our tool therefore discovered a total of 22 balance disorders (BD), 75 inheritance disorders (ID), 11 skip empty string literal (SESL), 5 type deduction overflows (TDO), 80 uninitialised structs (US), 382 hidden state updates (HSU), 14 hidden transfers (HT) and finally 101 straw man contracts (SMC).
While many contracts were found to be HSU, SMC and US honeypots, only a small number were found to be TDO honeypots.

\subsection{Validation}

In order to confirm the correctness of \textsc{HoneyBadger}, we performed a manual inspection of the source code of the contracts that have been flagged as honeypots.
We were able to collect through Etherscan the source code for 323 (70\%) of the flagged contracts.~We verified the flagged contracts by manually scanning the source code for characteristics of the detected honeypot technique.
For example, in case a contract has been flagged as a balance disorder, we checked whether the source code contains a function that transfers the contract's balance to the caller if and only if the value sent to the function is greater than or equal to the contract's balance.

\noindent
\newline
Table~\ref{tbl:positives_validation} summarises our manual verification in terms of true positives (TP), false positives (FP) and precision $p$, where $p$ is computed as $p = TP/(TP + FP)$. 
A true positive means that the contract is indeed a honeypot with respect to the reported technique and a false positive means that the contract is \emph{not} a honeypot with respect to the reported technique.
Overall our tool shows a very high precision and a very low false positive rate.
Our tool achieves a false positive rate of 0\% for 5 out of the 8 analysed honeypot techniques.
For the remaining 3 techniques, our tool achieves a decent false positive rate, where the highest false positive rate is roughly 18\% for the detection of hidden state updates, followed by 
15\% false positive rate for the detection of inheritance disorder and finally 12\% false positive rate for the detection of straw man contracts.

\begin{table}
\centering
\begin{small}
    \begin{tabular}{@{} cl*{10}c @{}}
        & & \rot{Balance Disorder} & \rot{Inheritance Disorder} 
        & \rot{Skip Empty String Literal} & \rot{Type Deduction Overflow}  
        & \rot{Uninitialised Struct} & \rot{Hidden State Update} & \rot{Hidden Transfer} 
        & \rot{Straw Man Contract} \\
        \cmidrule[1pt]{2-10}
        & TP & 20 & 41 & 9 & 4 & 32 & 134 & 12 & 30 & & \\
        & FP &   0 &   7 & 0 & 0 &   0 &   30 &   0 &   4 & & \\
        \cmidrule[1pt]{2-10}
        & $p$ & 100 & 85 & 100 & 100 & 100 & 82 & 100 & 88 & & \\
    \end{tabular}
    \caption{Number of true positives (TP), false positives (FP) and precision $p$ (in \%) per detected honeypot technique for contracts with source code.}
    \label{tbl:positives_validation}
\end{small}
\end{table}

\section{Analysis}
\label{sec:analysis}

In this section, we analyse the true positives obtained in Section~\ref{sec:evaluation}, in order to acquire insights on the \emph{effectiveness}, \emph{liveness}, \emph{behaviour}, \emph{diversity} and \emph{profitability} of honeypots.

    
\subsection{Methodology}

We crawled all the transactions of the 282 true positives using Etherchain's\footnote{https~://www.etherchain.org/} API, in order to collect various information about the honeypots, such as the amount of spent and received ether per address, the deployment date and the balance.~Afterwards, we used simple heuristics to label every address as either an \emph{attacker} or a \emph{victim}.
An address is labeled as an attacker if it either:~1)~created the honeypot;~2)~was the first address to send ether to the honeypot; or~3)~received more ether than it actually spent on the honeypot.~An address is labeled as a victim if it has not been labeled as an attacker and if it received less ether than it actually spent on the honeypot. 
Finally, using this information we were able to tell if a honeypot, was either \emph{successful}, \emph{aborted} or still \emph{active}.~A honeypot is marked as successful if a victim has been detected, as aborted if the balance is zero and no victim has been detected or as active if the balance is larger than zero and no victim has been detected.

\subsection{Results}

\paragraph{Effectiveness.}

\begin{figure}
\centering
\includegraphics[scale=0.7]{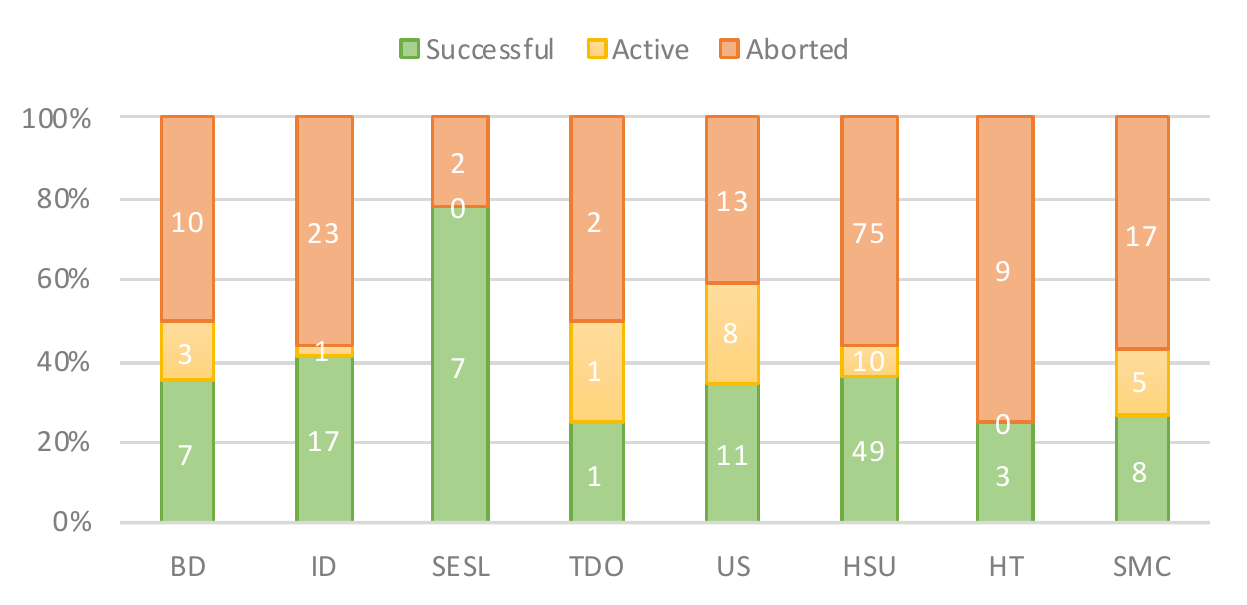}
\caption{Number of successful, active and aborted honeypots per honeypot technique.}
\label{fig:effectiveness}
\end{figure}

\begin{figure}
\centering
\includegraphics[scale=0.6]{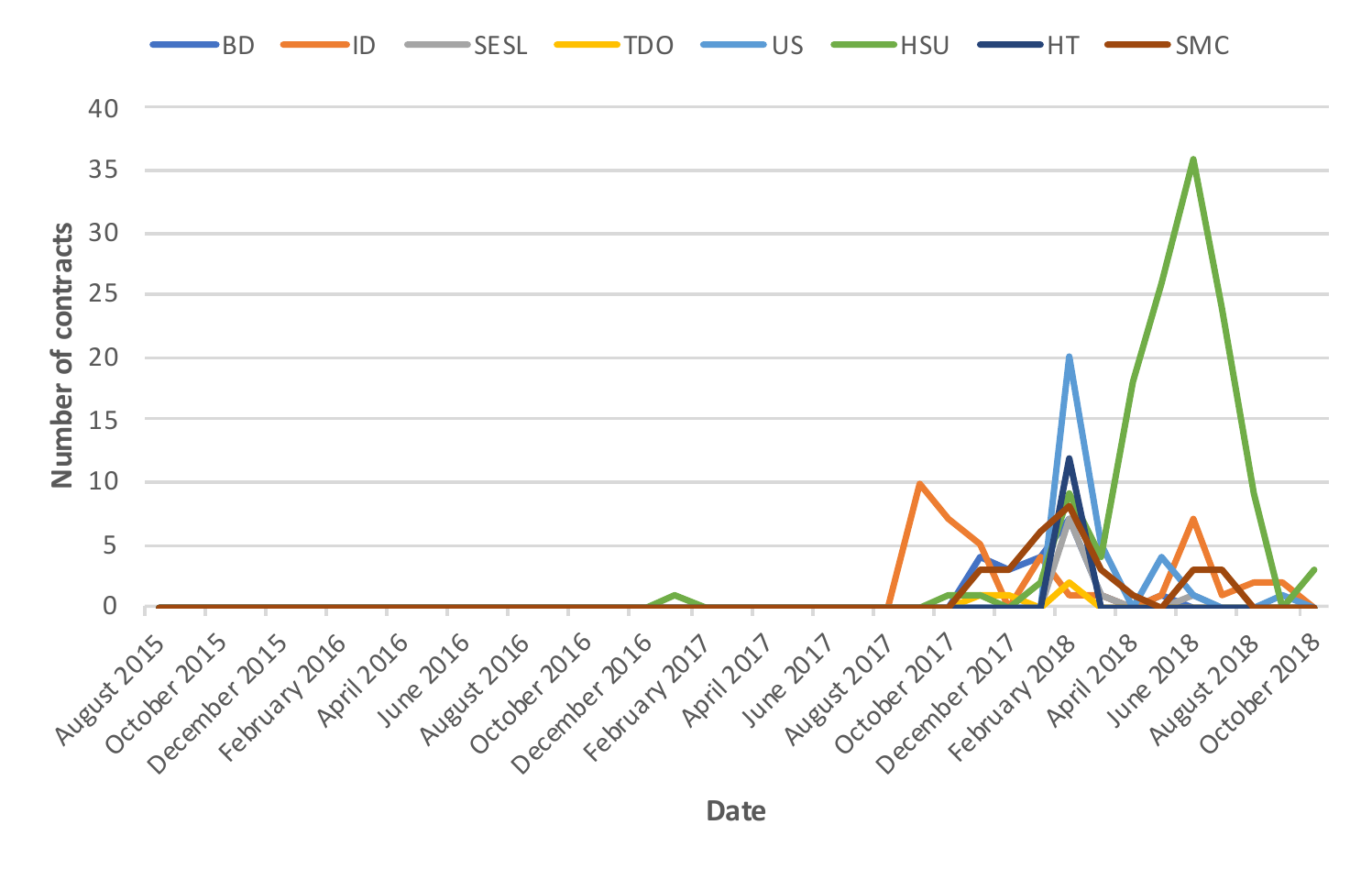}
\caption{Number of monthly deployed honeypots per honeypot technique.}
\label{fig:deployments}
\end{figure}

Figure~\ref{fig:effectiveness} shows the number of successful, aborted and active honeypots per honeypot technique.
Our results show that \emph{skip empty string literal} is the most effective honeypot technique with roughly 78\% success rate, whereas \emph{hidden transfer} is the least effective technique with solely 33\% success rate.
The overall success rate of honeypots seems rather low with roughly 37\%, whereas the overall abortion rate seems quite high with about 54\%. 
At the time of writing, solely 10\% of the analysed honeypots are still active.
Figure~\ref{fig:deployments} illustrates the number of monthly deployed honeypots per honeypot technique.
The very first honeypot technique that has been deployed was a hidden state update in January 2017.
February 2018 has been the peak in terms of honeypots being deployed, with a total of 66.
The highest number of monthly honeypots that have been deployed per technique are hidden state updates with a total of 36 in June 2018.
7 honeypots have been deployed on average per month.
In our analysis, the quickest first attempt of exploitation happened just 7 minutes and 37 seconds after a honeypot had been deployed, whereas the longest happened not until 142 days after deployment.
A honeypot takes an average of 9 days and a median of 16 hours before it gets exploited.
Interestingly, most honeypots (roughly 55\%) are exploited during the first 24 hours after being deployed.
 
\paragraph{Liveness.}

We define the lifespan of a honeypot as the period of time between the deployment of a honeypot and the moment when a honeypot was aborted.
We found that the shortest lifespan of a honeypot was 5 minutes and 25 seconds and the longest lifespan was about 322 days.
The average lifespan of a honeypot is roughly 28 days, whereas the median is roughly 3 days.
However, in around 32\% of the cases the lifespan of a honeypot is solely 1 day.~We also analysed how long an attacker keeps the funds inside a honeypot, by measuring the period of time between the first attempt of exploitation by a victim and the withdrawal of all the funds by the attacker.
The shortest period was just 4 minutes and 28 seconds after a victim fell for the honeypot.
The longest period was roughly 100 days.
On average attackers withdraw all their funds within 7 days after a victim fell for the honeypot.
However, in most cases the attackers keep the funds in the honeypot for a maximum of 1 day.
Interestingly, only 37 out of 282 honeypots got destroyed, where destroyed means that the attacker called a function within the honeypot that calls the \textsc{SELFDESTRUCT} opcode.
In other words, 171 honeypots are in some kind of ``zombie'' state, where they are still alive (i.e.~not destroyed), but not active (i.e.~their balance is zero). 
Analysing the 37 destroyed honeypots, we found that 19 got destroyed after being successful and 18 after never having been successful.

 \paragraph{Behaviour.}

\begin{figure}
\centering
\includegraphics[scale=0.4]{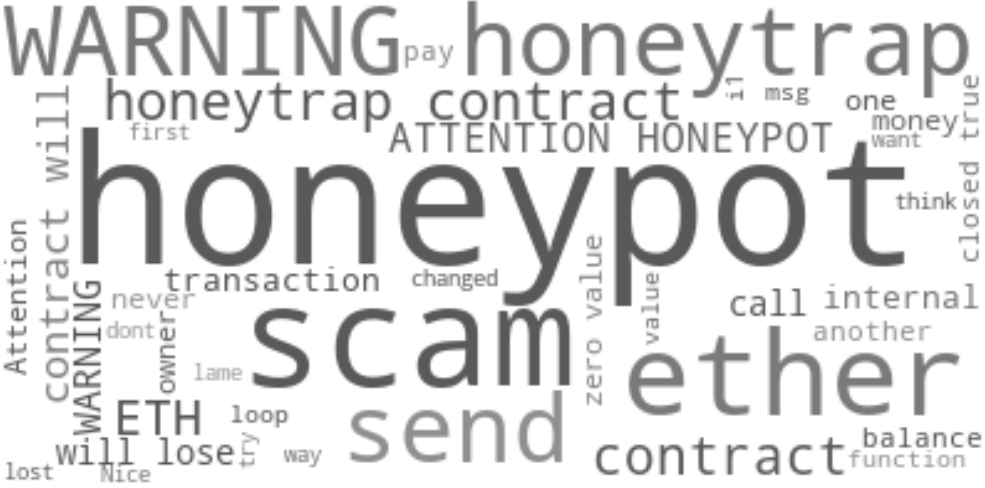}
\caption{A word cloud generated from the comments on Etherscan.}
\label{fig:etherscan_wordcloud}
\end{figure}

Our methodology classified a total of 240 addresses as victims.
In 71\% of the cases a honeypot managed to trap solely one victim.
In one case though, 97 victims have been trapped by just a single honeypot.
Interestingly, 8 out of the 240 addresses fell for more than one honeypot, where one address even became a victim to four different honeypots.
We also found that 53 attackers deployed at least two honeypots, whereas a sole attacker deployed eight different honeypots.
It is worth noting that 42 of the 53 attackers simply deployed copies of one particular honeypot type, whereas the remaining 11 deployed honeypots of varying types.
87 out of the 282 detected and manually confirmed honeypots (about 31\%) contained comments on Etherscan.
We manually analysed these comments and found that the majority of the comments were indeed warnings stating that the contract might be a honeypot. 
Moreover, Figure~\ref{fig:etherscan_wordcloud} shows that the term ``honeypot'' is the most prevalent term used by the community to describe this type of smart contracts.
Surprisingly, 20 out of the 87 commented honeypots were successful.
16 were successful before a comment had been placed and 4 have been successful even after a comment had been placed.
Interestingly, 21 honeypots aborted after a comment was placed.
The quickest abort was performed just 33 minutes and 57 seconds after the comment, whereas the longest abort was performed 37 days after the comment. 
Finally, attackers took an average of 6 days and a median of 22 hours to abort their honeypot after a user had placed a comment.

\paragraph{Diversity.}

We used the normalised Levenshtein distance~\cite{yujian2007normalized} to measure the similarity of the bytecode between the individual instances of a particular honeypot technique. 
Table~\ref{tbl:diversity} outlines the similarity in terms of minimum, maximum, mean and mode per honeypot technique.
We observe that for almost every technique, except TDO, the bytecode similarity varies tremendously.
For example, in case of hidden state update honeypots, we measure a minimum similarity of 11\% and a maximum similarity of 98\%. This indicates that even though two honeypots share the same technique, their bytecode might still be very diverse. 

\begin{table}[t]
\centering
\begin{footnotesize}
    \begin{tabular}{l*{8}c}
        \cmidrule[1pt]{1-9}
        & \textbf{BD} & \textbf{ID} & \textbf{SESL} & \textbf{TDO} & \textbf{US} & \textbf{HSU} & \textbf{HT} & \textbf{SMC} \\
        \cmidrule[0.5pt]{1-9}
        Min.   & 27 & 14 & 22 & 88 & 25 & 11 & 28 & 26 \\
        Max.  & 97 & 96 & 98 & 95 & 98 & 98 & 98 & 98 \\
        Mean & 50 & 40 & 47 & 90 & 52 & 49 & 71 & 53 \\
        Mode & 35 & 35 & 28 & 89 & 45 & 36 & 95 & 49 \\
        \cmidrule[1pt]{1-9}
    \end{tabular}
\caption{Bytecode similarity (in \%) per honeypot technique.}
\label{tbl:diversity}
\end{footnotesize}
\end{table}

\paragraph{Profitability.}

\begin{table}[b]
\centering
\begin{footnotesize}
    \begin{tabular}{l*{6}c}
        \cmidrule[1pt]{1-7}
        & \textbf{Min.} & \textbf{Max.} & \textbf{Mean} & \textbf{Mode} & \textbf{Median} & \textbf{Sum} \\
        \cmidrule[0.5pt]{1-7}
        BD & 0.01 & 1.13 & 0.5 & 0.11 & 0.11 & 3.5 \\
        ID & 0.004 & 6.41 & 1.06 & 0.1 & 0.33 & 17.02 \\
        SESL & 0.584 & 4.24 & 1.59 & 1.0 & 1.23 & 9.57 \\
        TDO & - & - & - & - & - & - \\
        US & 0.009 & 1.1 & 0.46 & 0.1 & 0.38 & 6.44 \\ 
        HSU & 0.00002 & 11.96 & 1.44 & 0.1 & 1.02 & 171.22 \\
        HT & 1.009 & 1.1 & 1.05 & 1.0 & 1.05 & 2.11 \\
        SMC & 0.399 & 4.94 & 1.76 & 2.0 & 1.99 & 47.39 \\
        \cmidrule[0.5pt]{1-7}
        Overall & 0.00002 & 11.96 & 1.35 & 1.0 & 1.01 & 257.25 \\
        \cmidrule[1pt]{1-7}
    \end{tabular}
\caption{Statistics on the profitability of each honeypot technique in ether.}
\label{tbl:profitability}
\end{footnotesize}
\end{table}

Table~\ref{tbl:profitability} lists the profitability per honeypot technique.
The profitability is computed as $\mathit{received\;amount}$ - $(\mathit{spent\;amount}$ + $\mathit{transaction\;fees})$.
No values are provided for TDO, because for the single true positive that we analysed, the transaction fees spent by the attacker were higher than the amount that the attacker gained from the victim.
The smallest and largest profit were made using a hidden state update honeypot, with 0.00002 ether being the smallest and 11.96 ether being the largest.
The most profitable honeypots are straw man contract honeypots, with an average value of 1.76 ether, whereas the least profitable honeypots are uninitialised struct honeypots, with an average value of 0.46 ether.
A total profit of 257.25 ether has been made through honeypots, of which 171.22 ether were solely made through hidden state update honeypots. 
However, the exchange rate of cryptocurrencies is very volatile and thus their value in USD may vary greatly on a day-to-day basis.
For example, although 11.96 ether is the largest profit made in ether, its actual value in USD was solely 500 at the point of withdrawal. 
Thus, we found that the largest profit in terms of USD, was actually a honeypot with 3.10987 ether, as it was worth 2,609 USD at the time of withdrawal.
Applying this method across the 282 honeypots, results in a total profit of 90,118 USD.

\section{Discussion}

In this section we summarise the key insights gained through our analysis and we discuss the ethical considerations as well as the challenges and limitations of our work.

\subsection{Honeypot Insights}


Although honeypots are capable of trapping multiple users, we have found that most honeypots managed to take the funds of only one victim.
This indicates that users potentially look at the transactions of other users before they submit theirs.
Moreover, the low success rate of honeypots with comments, suggests that users also check the comments on Etherscan before submitting any funds.
We also found that the bytecode of honeypots can be vastly different even if using the same honeypot technique.
This suggests that the usage of signature-based detection methods would be rather ineffective.
\textsc{HoneyBadger} is capable of recognising a variety of implementations, as it specifically targets the functional characteristics of each honeypot technique.
More than half of the honeypots were successful within the first 24 hours. 
This suggests that honeypots become less effective the older they become.
This is interesting, as it means that users seem to target rather recently deployed honeypots than older ones.
We also note that most honeypot creators withdraw their loot within 24 hours or abort their honeypots if they are not successful within the first 24 hours.
We therefore conclude that honeypots have in general a short lifespan and only a small fraction remain active for a period longer than one day.

\subsection{Challenges and Limitations}

The amount of smart contracts with source code available is rather small. 
At the time of writing, there are only 50,000 contracts with source code available on Etherscan.~This highlights the necessity of being able to detect honeypots at the bytecode level.
Unfortunately, this turns out to be extremely challenging when detecting certain honeypot techniques.
For example, while detecting inheritance disorder at the source code level is rather trivial, detecting it at the bytecode level is rather difficult since all information about the inheritance is lost during compilation and not available anymore at the bytecode level.
The fact that certain information is solely available at the source code level and not at the bytecode level, obliges us to make use of other less precise information that is available in the bytecode in order to detect honeypot techniques such as inheritance disorder.
However, as Section~\ref{sec:evaluation} shows, this approach reduces the precision of our detection and introduces some false positives.
Finally, another limitation of our tool is that it is currently limited to the detection of the eight honeypot techniques described in this paper.
Thus other honeypot techniques are not detected.
Nevertheless, we designed \textsc{HoneyBadger} with modularity in mind, such that one can easily extend the honeypot analysis component with new heuristics in order to detect more honeypot techniques. 

\subsection{Ethical Considerations}

In general, honeypots have two participants, the creator of the honeypot, and the user whose funds are trapped by the honeypot.
However, the ethical intentions of both participants are not always clear.
For instance, a honeypot creator might deploy a honeypot with the intention to scam users and make profit.
In this case we clearly have a malicious intention.
However, one could also argue that a honeypot creator is just attempting to punish users that behave maliciously.
Similarly, the intentions of a honeypot user can either be malicious or benign.
For example, if a user tries to intentionally exploit a reentrancy vulnerability, then he or she needs to be knowledgable and mischievous enough to prepare and attempt the attack, and thus clearly showing malicious behaviour.
However, if we take the example of an uninitialised struct honeypot that is disguised as a simple lottery,
then we might have the case of a benign user who loses his funds under the assumption that he or she is participating in a fair lottery.
Thus, both honeypot creators and users cannot always be clearly classified as either malicious or benign, this depends on the case at hand.
Nevertheless, we are aware that our methodology may serve malicious attackers to protect themselves from other malicious attackers.
However, with \textsc{HoneyBadger}, we hope to raise the awareness of honeypots and save benign users from potential financial losses.


\section{Related Work}

Honeypots are a new type of fraud that combine security issues with scams.
They either rely on the blockchain itself or on related services such as Etherscan.
With growing interest within the blockchain community, they have been discussed online \cite{honeypot_analysis, hacking_the_hackers, dissecting_honeypots} and collected within public user repositories \cite{solidity-vulnerable-honeypots, smart-contract-honeypots}.
Frauds and security issues are nothing new within the blockchain ecosystem.
Blockchains have been used for money laundering \cite{moser2013inquiry} and been the target of several scams \cite{vasek2015there}, including mining scams, wallet scams and Ponzi schemes, which are further discussed in \cite{bartoletti2018data, vasek2018analyzing}.
In particular, smart contracts have been shown to contain security issues \cite{Atzei2017}.
Although not performed directly on the blockchain, exchanges have also been the target of fraud \cite{moore2013beware}.

Several different methods have been proposed to discover fraud as well as security issues.
Manual analysis is performed on publicly available source code to detect Ponzi schemes \cite{bartoletti2017dissecting}.
\cite{zhou2018erays} introduces \textsc{Erays}, a tool that aims to produce easy to analyse pseudocode from bytecode where the source code is not available.
However, manual analysis is particularly laborious, especially considering the number of contracts on the blockchain.
Machine learning has been used to detect Ponzi schemes \cite{chen2018detecting} and to find vulnerabilities \cite{tann2018towards}.
The latter relies on \cite{nikolic2018finding} to obtain a ground truth of vulnerable smart contracts for training their model.
Fuzzing techniques have been employed to detect security vulnerabilities in smart contracts \cite{jiang2018contractfuzzer} and in combination with symbolic execution to discover issues related to the ordering of events or function calls \cite{kolluri2018exploiting}.
However, fuzzing often fails to create inputs to enter specific execution paths and therefore might ignore them \cite{vanhoef2018symbolic}.
Static analysis has been used to find security \cite{brent2018vandal, tsankov2018securify, tikhomirov2018smartcheck} and gas-focused \cite{grech2018madmax} vulnerabilities in smart contracts.
\cite{brent2018vandal} requires manual interaction, while \cite{tsankov2018securify} requires both the definition of violation and compliance patterns.
\cite{tikhomirov2018smartcheck} requires Solidity code and therefore cannot be used to analyse the large majority of the smart contracts deployed on the Ethereum blockchain.
\cite{grech2018madmax} considers gas-related issues which is not necessary for the purpose of this work.
In order to use formal verification, smart contracts can, to some extent, be translated from source code or bytecode into F* \cite{Bhargavan2016, grishchenko2018semantic} where the verification can more easily be performed.
Other work operates on high-level source code available for Ethereum or Hyperledger \cite{kalra2018zeus}.
\cite{hirai2017defining, coq} propose a formal definition of the EVM, that is extended in \cite{amani2018towards} towards more automated smart contract verification and the consideration of gas.
Formal verification often requires (incomplete) translations into other languages or manual user interaction (e.g.: \cite{why3}).
Both of these reasons make formal verification unsuitable to be used on a large number of contracts, as it is required in this work.

Symbolic execution has been used on smart contracts to detect common \cite{manticore, mueller2018, Luu2016, torres2018osiris} vulnerabilities.
This technique also allows to find specific kinds of misbehaving contracts \cite{nikolic2018finding}.
It can further provide values that can serve to generate automated exploits that trigger vulnerabilities \cite{krupp2018teether}.
The same technique is used in this paper.
Symbolic execution has the advantage of being capable to reason about all possible execution paths and states in a smart contract.
This allows for the implementation of precise heuristics while achieving a low false positive rate.
Another advantage is that symbolic execution can be applied directly to bytecode, thus making it well suited for our purpose of analysing more than 2 million smart contracts for which source code is largely not available.
The disadvantage is the large number of possible paths that need to be analysed.
However, in the case of smart contracts this is not an issue, as most are not very complex and very short.
Moreover, smart contract bytecode cannot grow arbitrarily large due to the gas limit enforced by the Ethereum blockchain.

To the best of the authors' knowledge, this paper is the first to consider and discuss honeypot smart contracts, a new type of fraud, and to propose a taxonomy as well as an automated tool using symbolic execution for their detection.

\section{Conclusion}

In this work, we investigated an emerging new type of fraud in Ethereum: \emph{honeypots}.
We presented a taxonomy of honeypot techniques and introduced a methodology that uses symbolic execution and heuristics for the automated detection of honeypots.
We showed that \textsc{HoneyBadger} can effectively detect honeypots in the wild with a very low false positive rate.~In a large-scale analysis of 151,935 unique Ethereum smart contracts, \textsc{HoneyBadger} identified 460 honeypots.
Moreover, an analysis on the transactions performed by a subset of 282 honeypots, revealed that 240 users already became victims of honeypots and that attackers already made more than 90,000 USD profit with honeypots.
It is worth noting that these numbers solely provide a lower bound and thus might only reflect the tip of the iceberg. 
Nonetheless, tools such as \textsc{HoneyBadger} may already help users in detecting honeypots before they can cause any harm.
In future work, we plan to further generalise our detection mechanism through the use of machine learning techniques. 
We also plan to extend our analysis with a larger subset and eventually detect new honeypots by looking at other contracts that are linked to the newly discovered honeypot contracts. 

\section*{Acknowledgments}

We would like to thank Hugo Jonker and Sjouke Mauw as well as the anonymous reviewers for their valuable feedback and comments.
The experiments presented in this paper were carried out using the HPC facilities of the University of Luxembourg~\cite{VBCG_HPCS14}
{\small -- see~\url{https://hpc.uni.lu}}. 
This work is partly supported by the Luxembourg National Research Fund (FNR) under grant 13192291.

{\normalsize \bibliographystyle{plain}
\bibliography{references}}

\begin{thebibliography}{10}

\bibitem{amani2018towards}
Sidney Amani, Myriam B{\'e}gel, Maksym Bortin, and Mark Staples.
\newblock Towards verifying ethereum smart contract bytecode in isabelle/hol.
\newblock {\em CPP. ACM. To appear}, 2018.

\bibitem{Atzei2017}
Nicola Atzei, Massimo Bartoletti, and Tiziana Cimoli.
\newblock {A Survey of Attacks on Ethereum Smart Contracts (SoK)}.
\newblock In {\em Proceedings of the 6th International Conference on Principles
  of Security and Trust - Volume 10204}, pages 164--186. Springer-Verlag New
  York, Inc., 2017.

\bibitem{bartoletti2017dissecting}
Massimo Bartoletti, Salvatore Carta, Tiziana Cimoli, and Roberto Saia.
\newblock Dissecting ponzi schemes on ethereum: identification, analysis, and
  impact.
\newblock {\em arXiv preprint arXiv:1703.03779}, 2017.

\bibitem{bartoletti2018data}
Massimo Bartoletti, Barbara Pes, and Sergio Serusi.
\newblock Data mining for detecting bitcoin ponzi schemes.
\newblock {\em arXiv preprint arXiv:1803.00646}, 2018.

\bibitem{Bhargavan2016}
Karthikeyan Bhargavan, Nikhil Swamy, Santiago Zanella-B{\'{e}}guelin, Antoine
  Delignat-Lavaud, C{\'{e}}dric Fournet, Anitha Gollamudi, Georges Gonthier,
  Nadim Kobeissi, Natalia Kulatova, Aseem Rastogi, and Thomas Sibut-Pinote.
\newblock {Formal Verification of Smart Contracts}.
\newblock In {\em Proceedings of the 2016 ACM Workshop on Programming Languages
  and Analysis for Security - PLAS'16}, pages 91--96, New York, New York, USA,
  2016. ACM Press.

\bibitem{bamboo}
Cornell Blockchain.
\newblock Bamboo: a language for morphing smart contracts, May 2018.
\newblock https://github.com/CornellBlockchain/bamboo.

\bibitem{brent2018vandal}
Lexi Brent, Anton Jurisevic, Michael Kong, Eric Liu, Francois Gauthier, Vincent
  Gramoli, Ralph Holz, and Bernhard Scholz.
\newblock Vandal: A scalable security analysis framework for smart contracts.
\newblock {\em arXiv preprint arXiv:1809.03981}, 2018.

\bibitem{chen2018detecting}
Weili Chen, Zibin Zheng, Jiahui Cui, Edith Ngai, Peilin Zheng, and Yuren Zhou.
\newblock Detecting ponzi schemes on ethereum: Towards healthier blockchain
  technology.
\newblock In {\em Proceedings of the 2018 World Wide Web Conference on World
  Wide Web}, pages 1409--1418. International World Wide Web Conferences
  Steering Committee, 2018.

\bibitem{coinmarketcap}
{CoinMarketCap}.
\newblock {Ethereum (ETH) price, charts, market cap, and other metrics |
  CoinMarketCap}, January 2018.
\newblock https://coinmarketcap.com/currencies/ethereum/.

\bibitem{de2008z3}
Leonardo De~Moura and Nikolaj Bj{\o}rner.
\newblock Z3: An efficient smt solver.
\newblock In {\em International conference on Tools and Algorithms for the
  Construction and Analysis of Systems}, pages 337--340. Springer, 2008.

\bibitem{grech2018madmax}
Neville Grech, Michael Kong, Anton Jurisevic, Lexi Brent, Bernhard Scholz, and
  Yannis Smaragdakis.
\newblock Madmax: surviving out-of-gas conditions in ethereum smart contracts.
\newblock {\em Proceedings of the ACM on Programming Languages}, 2(OOPSLA):116,
  2018.

\bibitem{grishchenko2018semantic}
Ilya Grishchenko, Matteo Maffei, and Clara Schneidewind.
\newblock A semantic framework for the security analysis of ethereum smart
  contracts.
\newblock In {\em International Conference on Principles of Security and
  Trust}, pages 243--269. Springer, 2018.

\bibitem{hirai2017defining}
Yoichi Hirai.
\newblock Defining the ethereum virtual machine for interactive theorem
  provers.
\newblock In {\em International Conference on Financial Cryptography and Data
  Security}, pages 520--535. Springer, 2017.

\bibitem{coq}
Yoichi Hirai.
\newblock Ethereum virtual machine for coq (v0.0.2), June 2017.
\newblock
  https://medium.com/@pirapira/ethereum-virtual-machine-for-coq-v0-0-2-d2568e068b18.

\bibitem{jiang2018contractfuzzer}
Bo~Jiang, Ye~Liu, and W.~K. Chan.
\newblock Contractfuzzer: Fuzzing smart contracts for vulnerability detection.
\newblock In {\em Proceedings of the 33rd ACM/IEEE International Conference on
  Automated Software Engineering}, ASE 2018, pages 259--269, New York, NY, USA,
  2018. ACM.

\bibitem{kalra2018zeus}
Sukrit Kalra, Seep Goel, Mohan Dhawan, and Subodh Sharma.
\newblock Zeus: Analyzing safety of smart contracts.
\newblock In {\em NDSS}, 2018.

\bibitem{kolluri2018exploiting}
Aashish Kolluri, Ivica Nikolic, Ilya Sergey, Aquinas Hobor, and Prateek Saxena.
\newblock Exploiting the laws of order in smart contracts.
\newblock {\em arXiv preprint arXiv:1810.11605}, 2018.

\bibitem{krupp2018teether}
Johannes Krupp and Christian Rossow.
\newblock teether: Gnawing at ethereum to automatically exploit smart
  contracts.
\newblock In {\em 27th {USENIX} Security Symposium ({USENIX} Security 18)},
  pages 1317--1333, 2018.

\bibitem{lll}
LLL.
\newblock Ethereum low-level lisp-like language, January 2019.
\newblock https://lll-docs.readthedocs.io/en/latest/lll\_introduction.html.

\bibitem{oyente}
Loi Luu.
\newblock {Oyente - An Analysis Tool for Smart Contracts v0.2.7
  (Commonwealth)}, February 2017.
\newblock https://github.com/melonproject/oyente.

\bibitem{Luu2016}
Loi Luu, Duc-Hiep Chu, Hrishi Olickel, Prateek Saxena, and Aquinas Hobor.
\newblock Making smart contracts smarter.
\newblock In {\em Proceedings of the 2016 ACM SIGSAC Conference on Computer and
  Communications Security}, CCS '16, pages 254--269, New York, NY, USA, 2016.
  ACM.

\bibitem{solidity-vulnerable-honeypots}
misterch0c.
\newblock Solidity vulnerable honeypots, April 2018.
\newblock
  https://github.com/misterch0c/Solidlity-Vulnerable/tree/master/honeypots.

\bibitem{moore2013beware}
Tyler Moore and Nicolas Christin.
\newblock Beware the middleman: Empirical analysis of bitcoin-exchange risk.
\newblock In {\em International Conference on Financial Cryptography and Data
  Security}, pages 25--33. Springer, 2013.

\bibitem{moser2013inquiry}
Malte Moser, Rainer Bohme, and Dominic Breuker.
\newblock An inquiry into money laundering tools in the bitcoin ecosystem.
\newblock In {\em eCrime Researchers Summit (eCRS), 2013}, pages 1--14. IEEE,
  2013.

\bibitem{mueller2018}
Bernhard Mueller.
\newblock Smashing ethereum smart contracts for fun and real profit.
\newblock In {\em 9th annual HITB Security Conference}, 2018.

\bibitem{nakamoto2008bitcoin}
Satoshi Nakamoto.
\newblock Bitcoin: A peer-to-peer electronic cash system.
\newblock {\em Cryptography Mailing list at https://metzdowd.com}, 03 2009.

\bibitem{nikolic2018finding}
Ivica Nikolic, Aashish Kolluri, Ilya Sergey, Prateek Saxena, and Aquinas Hobor.
\newblock Finding the greedy, prodigal, and suicidal contracts at scale.
\newblock {\em arXiv preprint arXiv:1802.06038}, 2018.

\bibitem{manticore}
Trail of~Bits.
\newblock Manticore - symbolic execution tool, jun 2018.
\newblock https://github.com/trailofbits/manticore.

\bibitem{parity_hack}
Sergey Petrov.
\newblock Another parity wallet hack explained, nov 2017.
\newblock
  https://medium.com/@Pr0Ger/another-parity-wallet-hack-explained-847ca46a2e1c.

\bibitem{why3}
Christian Reitwiessner.
\newblock Formal verification for solidity contracts, June 2018.
\newblock
  https://forum.ethereum.org/discussion/3779/formal-verification-for-solidity-contracts.

\bibitem{honeypot_analysis}
Josep Sanjuas.
\newblock An analysis of a couple ethereum honeypot contracts, December 2018.
\newblock
  https://medium.com/coinmonks/an-analysis-of-a-couple-ethereum-honeypot-contracts-5c07c95b0a8d.

\bibitem{hacking_the_hackers}
Alex Sherbachev.
\newblock Hacking the hackers: Honeypots on ethereum network, December 2018.
\newblock
  https://hackernoon.com/hacking-the-hackers-honeypots-on-ethereum-network-5baa35a13577.

\bibitem{dissecting_honeypots}
Alex Sherbuck.
\newblock Dissecting an ethereum honey pot, December 2018.
\newblock
  https://medium.com/coinmonks/dissecting-an-ethereum-honey-pot-7102d7def5e0.

\bibitem{dao_hack}
David Siegel.
\newblock Understanding the dao attack, jun 2016.
\newblock https://www.coindesk.com/understanding-dao-hack-journalists/.

\bibitem{szabo1997}
Nick Szabo.
\newblock Formalizing and securing relationships on public networks.
\newblock {\em First Monday}, 2(9), 1997.

\bibitem{tann2018towards}
A~Tann, Xing~Jie Han, Sourav~Sen Gupta, and Yew-Soon Ong.
\newblock Towards safer smart contracts: A sequence learning approach to
  detecting vulnerabilities.
\newblock {\em arXiv preprint arXiv:1811.06632}, 2018.

\bibitem{tikhomirov2018smartcheck}
S.~Tikhomirov, E.~Voskresenskaya, I.~Ivanitskiy, R.~Takhaviev, E.~Marchenko,
  and Y.~Alexandrov.
\newblock Smartcheck: Static analysis of ethereum smart contracts.
\newblock In {\em 2018 IEEE/ACM 1st International Workshop on Emerging Trends
  in Software Engineering for Blockchain (WETSEB)}, pages 9--16, May 2018.

\bibitem{torres2018osiris}
Christof~Ferreira Torres, Julian Sch\"{u}tte, and Radu State.
\newblock Osiris: Hunting for integer bugs in ethereum smart contracts.
\newblock In {\em Proceedings of the 34th Annual Computer Security Applications
  Conference}, ACSAC '18, pages 664--676, New York, NY, USA, 2018. ACM.

\bibitem{tsankov2018securify}
Petar Tsankov, Andrei Dan, Dana~Drachsler Cohen, Arthur Gervais, Florian
  Buenzli, and Martin Vechev.
\newblock Securify: Practical security analysis of smart contracts.
\newblock {\em arXiv preprint arXiv:1806.01143}, 2018.

\bibitem{vanhoef2018symbolic}
Mathy Vanhoef and Frank Piessens.
\newblock Symbolic execution of security protocol implementations: Handling
  cryptographic primitives.
\newblock In {\em 12th {USENIX} Workshop on Offensive Technologies ({WOOT}
  18)}, Baltimore, MD, 2018. {USENIX} Association.

\bibitem{VBCG_HPCS14}
S.~Varrette, P.~Bouvry, H.~Cartiaux, and F.~Georgatos.
\newblock Management of an academic hpc cluster: The ul experience.
\newblock In {\em Proc. of the 2014 Intl. Conf. on High Performance Computing
  \& Simulation (HPCS 2014)}, pages 959--967, Bologna, Italy, July 2014. IEEE.

\bibitem{vasek2015there}
Marie Vasek and Tyler Moore.
\newblock There’s no free lunch, even using bitcoin: Tracking the popularity
  and profits of virtual currency scams.
\newblock In {\em International conference on financial cryptography and data
  security}, pages 44--61. Springer, 2015.

\bibitem{vasek2018analyzing}
Marie Vasek and Tyler Moore.
\newblock Analyzing the bitcoin ponzi scheme ecosystem.
\newblock In {\em Bitcoin Workshop}, 2018.

\bibitem{vyper}
Vyper.
\newblock Pythonic smart contract language for the evm, January 2019.
\newblock https://github.com/ethereum/vyper.

\bibitem{smart-contract-honeypots}
Gerhard Wagner.
\newblock Smart contract honeypots, April 2018.
\newblock https://github.com/thec00n/smart-contract-honeypots.

\bibitem{wood2014ethereum}
Gavin Wood.
\newblock Ethereum: A secure decentralised generalised transaction ledger.
\newblock {\em Ethereum Project Yellow Paper}, 151:1--32, 2014.

\bibitem{solidity}
Gavin Wood.
\newblock Solidity 0.5.1 documentation, December 2018.
\newblock https://solidity.readthedocs.io/en/v0.5.1/.

\bibitem{yujian2007normalized}
Li~Yujian and Liu Bo.
\newblock A normalized levenshtein distance metric.
\newblock {\em IEEE transactions on pattern analysis and machine intelligence},
  29(6):1091--1095, 2007.

\bibitem{zhou2018erays}
Yi~Zhou, Deepak Kumar, Surya Bakshi, Joshua Mason, Andrew Miller, and Michael
  Bailey.
\newblock Erays: Reverse engineering ethereum's opaque smart contracts.
\newblock In {\em 27th {USENIX} Security Symposium ({USENIX} Security 18)},
  pages 1371--1385, 2018.

\end{thebibliography}

\appendix

\section{List of Honeypots}
\label{appendix:a}

Table~\ref{tbl:list_of_honeypots} presents the list of 24 honeypots that have been collected from public sources available on the Internet. 




\begin{table*}
\centering
\begin{normalsize}
\begin{tabular}{c c c}
\cmidrule[1pt]{1-3}
\textbf{Contract Name} & \textbf{Contract Address} &  \textbf{Technique} \\ \cmidrule[1pt]{1-3} 

\multicolumn{3}{c}{\textbf{Ethereum Virtual Machine}} \\ \cmidrule[0.5pt]{1-3}

MultiplicatorX3 & \href{https://etherscan.io/address/0x5aa88d2901c68fda244f1d0584400368d2c8e739#code}{0x5aa88d2901c68fda244f1d0584400368d2c8e739} & Balance Disorder \\ 

PinCodeEtherStorage & \href{https://etherscan.io/address/0x35c3034556b81132e682db2f879e6f30721b847c#code}{0x35c3034556b81132e682db2f879e6f30721b847c} & Balance Disorder \\ \cmidrule[0.5pt]{1-3} 

\multicolumn{3}{c}{\textbf{Solidity Compiler}} \\ \cmidrule[0.5pt]{1-3} 

TestBank & \href{https://etherscan.io/address/0x70c01853e4430cae353c9a7ae232a6a95f6cafd9#code}{0x70c01853e4430cae353c9a7ae232a6a95f6cafd9} & Inheritance Disorder \\ 

KingOfTheHill & \href{https://etherscan.io/address/0x4dc76cfc65b14b3fd83c8bc8b895482f3cbc150a#code}{0x4dc76cfc65b14b3fd83c8bc8b895482f3cbc150a} & Inheritance Disorder \\ 

RichestTakeAll & \href{https://etherscan.io/address/0xe65c53087e1a40b7c53b9a0ea3c2562ae2dfeb24#code}{0xe65c53087e1a40b7c53b9a0ea3c2562ae2dfeb24} & Inheritance Disorder \\ 

ICO\_Hold & \href{https://etherscan.io/address/0x4ba0d338a7c41cc12778e0a2fa6df2361e8d8465#code}{0x4ba0d338a7c41cc12778e0a2fa6df2361e8d8465} & Inheritance Disorder \\ 

TerrionFund & \href{https://etherscan.io/address/0x33685492a20234101b553d2a429ae8a6bf202e18#code}{0x33685492a20234101b553d2a429ae8a6bf202e18} & Inheritance Disorder \\

DividendDistributorv3 & \href{https://etherscan.io/address/0x858c9eaf3ace37d2bedb4a1eb6b8805ffe801bba#code}{0x858c9eaf3ace37d2bedb4a1eb6b8805ffe801bba} & Skip Empty String Literal \\

For\_Test & \href{https://etherscan.io/address/0x2ecf8d1f46dd3c2098de9352683444a0b69eb229#code}{0x2ecf8d1f46dd3c2098de9352683444a0b69eb229} & Type Deduction Overflow \\

Test1 & \href{https://etherscan.io/address/0x791d0463b8813b827807a36852e4778be01b704e#code}{0x791d0463b8813b827807a36852e4778be01b704e}  & Type Deduction Overflow \\ 

CryptoRoulette & \href{https://etherscan.io/address/0x94602b0e2512ddad62a935763bf1277c973b2758#code}{0x94602b0e2512ddad62a935763bf1277c973b2758} & Uninitialised Struct \\ 

OpenAddressLottery & \href{https://etherscan.io/address/0xd1915a2bcc4b77794d64c4e483e43444193373fa#code}{0xd1915a2bcc4b77794d64c4e483e43444193373fa} & Uninitialised Struct \\ 

GuessNumber & \href{https://etherscan.io/address/0x559cc6564ef51bd1ad9fbe752c9455cb6fb7feb1#code}{0x559cc6564ef51bd1ad9fbe752c9455cb6fb7feb1} & Uninitialised Struct \\ \cmidrule[0.5pt]{1-3}

\multicolumn{3}{c}{\textbf{Etherscan Blockchain Explorer}} \\ \cmidrule[0.5pt]{1-3} 

TestToken & \href{https://etherscan.io/address/0x3d8a10ce3228cb428cb56baa058d4432464ea25d#code}{0x3d8a10ce3228cb428cb56baa058d4432464ea25d} & Hidden Transfer \\

WhaleGiveaway1 & \href{https://etherscan.io/address/0x7a4349a749e59a5736efb7826ee3496a2dfd5489#code}{0x7a4349a749e59a5736efb7826ee3496a2dfd5489} & Hidden Transfer \\ 

Gift\_1\_ETH & \href{https://etherscan.io/address/0xd8993f49f372bb014fb088eabec95cfdc795cbf6#code}{0xd8993f49f372bb014fb088eabec95cfdc795cbf6} & Hidden State Update \\

NEW\_YEARS\_GIFT & \href{https://etherscan.io/address/0x13c547ff0888a0a876e6f1304eaefe9e6e06fc4b#code}{0x13c547ff0888a0a876e6f1304eaefe9e6e06fc4b} & Hidden State Update \\ 

G\_GAME & \href{https://etherscan.io/address/0x3caf97b4d97276d75185aaf1dcf3a2a8755afe27#code}{0x3caf97b4d97276d75185aaf1dcf3a2a8755afe27} & Hidden State Update \\ 

IFYKRYGE & \href{https://etherscan.io/address/0x1237b26652eebf1cb8f59e07e07101c0df4f60f6#code}{0x1237b26652eebf1cb8f59e07e07101c0df4f60f6}  & Hidden State Update \\ 

EtherBet & \href{https://etherscan.io/address/0x3c3f481950fa627bb9f39a04bccdc88f4130795b#code}{0x3c3f481950fa627bb9f39a04bccdc88f4130795b} & Hidden State Update \\ 

Private\_Bank & \href{https://etherscan.io/address/0xd116d1349c1382b0b302086a4e4219ae4f8634ff#code}{0xd116d1349c1382b0b302086a4e4219ae4f8634ff} & Straw Man Contract \\

firstTest  & \href{https://etherscan.io/address/0x42db5bfe8828f12f164586af8a992b3a7b038164#code}{0x42db5bfe8828f12f164586af8a992b3a7b038164} & Straw Man Contract  \\ 

TransferReg & \href{https://etherscan.io/address/0x62d5c4a317b93085697cfb1c775be4398df0678c#code}{0x62d5c4a317b93085697cfb1c775be4398df0678c} & Straw Man Contract  \\ 

testBank & \href{https://etherscan.io/address/0x477d1ee2f953a2f85dbecbcb371c2613809ea452#code}{0x477d1ee2f953a2f85dbecbcb371c2613809ea452} & Straw Man Contract  \\ 


\cmidrule[1pt]{1-3}

\end{tabular}
\end{normalsize}
\caption{List of publicly available honeypots on the Internet~\cite{smart-contract-honeypots, solidity-vulnerable-honeypots, hacking_the_hackers, honeypot_analysis, dissecting_honeypots}.}
\label{tbl:list_of_honeypots}
\end{table*}

\end{document}